\documentclass[twocolumn,prb,superscriptaddress,showpacs]{revtex4-1}
\pdfoutput=1
\usepackage{amsmath,amssymb}
\usepackage{graphicx}
\usepackage{verbatim}
\usepackage{amsfonts}
\usepackage{dcolumn}
\usepackage{bm}
\usepackage{color}
\usepackage[colorlinks,citecolor=blue]{hyperref}
\usepackage{subfigure}

\begin{document}

\title{\large \bf $U(1)\times U(1)$ symmetry protected topological Order  in Gutzwiller wave functions}

\author{Zheng-Xin Liu}
\affiliation{Institute for Advanced Study, Tsinghua University, Beijing, 100084, P. R. China}
\affiliation{Perimeter Institute for Theoretical Physics, Waterloo, Ontario, N2L 2Y5 Canada}
\author{Jia-Wei Mei}
\affiliation{Perimeter Institute for Theoretical Physics, Waterloo, Ontario, N2L 2Y5 Canada}
\author{Peng Ye}
\affiliation{Perimeter Institute for Theoretical Physics, Waterloo, Ontario, N2L 2Y5 Canada}
\author{Xiao-Gang Wen}
\affiliation{Department of Physics, Massachusetts Institute of Technology, Cambridge, Massachusetts 02139, USA}
\affiliation{Perimeter Institute for Theoretical Physics, Waterloo, Ontario, N2L 2Y5 Canada}
\affiliation{Institute for Advanced Study, Tsinghua University, Beijing, 100084, P. R. China}

\begin{abstract}

Gutzwiller projection is a way to construct many-body wave functions that could carry topological order or symmetry protected topological (SPT) order. However, an important issue is to determine whether or not a given Gutzwiller-projected wave functions (GWF) carries a non-trivial SPT order, and which SPT order is carried by the wavefunction. In this paper, we numerically study the SPT order in a spin $S = 1$  GWF on the Kagome lattice. Using the standard Monte Carlo method, we directly confirm that the GWF has (1) gapped bulk with short-range correlations, (2) a trivial topological order via  nondegenerate ground state, and zero topological entanglement entropy, (3) a non-trivial $U(1)\times U(1)$ SPT order via the Hall conductances of the protecting $U(1)\times U(1)$ symmetry,  and (4) symmetry protected gapless boundary. %To our knowledge it is the first 
This represents numerical evidence of continuous symmetry protected topological order in %2D 
two-dimensional 
Bosonic lattice systems.

\end{abstract}

\maketitle

\section{Introduction}

Topological order \cite{Wen89, WenNiu90, Wen1990} was introduced to describe
exotic quantum phases without symmetry breaking, such as fractional quantum
Hall states\cite{TsuiStormerGossard1982,Laughlin1983} or spin liquid
sates.\cite{Anderson1973,Anderson1987} Opposite to Landau's paradigm of
symmetry breaking orders,\cite{Landau37, GL5064} topologically ordered phases
can not be distinguished by local order parameters.  It was shown that
different topological orders differ by many-body
entanglement.\cite{ChenGuWen2010} From this point of view, long-range entangled
states are topologically ordered and are characterized by exotic properties,
such as degeneracy of ground states on a torus, fractional excitations,
non-zero topological entanglement entropy\cite{PreskillKitaev06_TEE,
LevinWen06_TEE}. On the other hand, a short range entangled state is trivial
and can be adiabatically connected to a direct product state. However, if the
system has a symmetry, the phase diagram will be enriched. Even short-range
entangled states can belong to different phases, called symmetry protected
topological (SPT) phases.\cite{GuWen2009,Pollmann2010} Haldane
phase\cite{HaldanePLA1983,HaldanePRL1983} and topological
insulators\cite{KM0502, BZ0602, MB0706, FKM0703, QHZ0824} are typical examples
of phases that contain SPT orders. If the symmetry of a bosonic system is
described by group $G$, then a large class of SPT phases in $d+1$-dimension can
be constructed via group cohomology $\mathcal H^{d+1}(G,U(1))$
\cite{ChenGuLiuWen2011} or through nonlinear sigma
models.\cite{ChenGuLiuWen2011,Xu13} In 2+1D, many SPT phases can also be
understood through Chern-Simons effective theory.\cite{LuVishwanath2012}
Similar to quantum Hall states and topological insulators, the boundary of a
2+1D SPT phase must by gapless if the symmetry is not broken. For continuous
symmetry groups such as
$U(1)$\cite{ChenGuLiuWen2011,LuVishwanath2012,ChenWen2012, SenthilLevin13,
RegnaultSenthil13} or $SO(3)$,\cite{LiuWen2012}  different SPT phases can be
distinguished by Hall conductance, which are quantized to 2.  We would like to remark
that, before the recent studies of symmetry protected short-range entangled
states with trivial topological order ({\it i.e.} the SPT states), some
progress was made on symmetry enriched long-range entangled states with
nontrivoal topological order, the so called symmetry enriched topological
states,\cite{W0213,KLW0834,KW0906,MR1315,HungWan2013,LuVishwanath2013} where
the ``fractionalized representation'' of the symmetry, carried by topological
excitations and described by projective symmetry
group,\cite{W0213,KLW0834,KW0906} played a key role.

Although it is believed that symmetry can enrich quantum phases of matter, it
lacks simple lattice models to realize these nontrivial phases in spatial
dimension higher than 1+1D. SPT phases for discrete symmetry groups were
understood quite well, since the ground state wave functions and exactly
solvable models (which are usually complicated and contain many-body
interactions) for nontrivial SPT phases can be
constructed.\cite{ChenLiuWen2011,LevinGu2012} It is more challenging to realize
continuous symmetry protected phases. A $U(1)$ symmetry protected nontrivial
phase was reported in a continuous bose model, \cite{RegnaultSenthil13} and
lattice models that may realize continuous (or combined) symmetry protected
topological phases were proposed.\cite{ LuLee12, YeWen13, LiuGuWen14, MeiWen14}
In Ref.~\onlinecite{YeWen13}, the authors proposed projective construction of $SU(2)$ or $SO(3)$ SPT states. And lattice model Hamiltonians that may possibly stabilize SPT states with continuous symmetries were designed recently. \cite{ LiuGuWen14, MeiWen14, Motrunich14}

Using Gutzwiller-projected wave functions (GWF), we can construct different kinds of SPT states. 
In the present paper, we will numerically study a spin-1 state on the Kagome lattice constructed by Gutzwiller projected Chern Bands, which was firstly proposed in Ref.~\onlinecite{LuLee12}. We will show that this state is a $U(1)\times U(1)$ SPT state, where the two $U(1)$ groups correspond to $\sum S_{z,i}$ conservation and $\sum S_{z,i}^2$ conservation, respectively. This SPT state has the following properties:
it is gapped without conventional long range spin order; it has unique ground state and zero topological entanglement entropy; it has non-zero spin Hall
conductance, the $U(1)\times U(1)$ charge is not fractionalized; the boundary is gapless if the symmetry is reserved but can be gapped out by the perturbations that break the symmetry. As a comparison, we also study a $S=1$ chiral spin liquid state\cite{Tu14SUN_1}
which is long-range entangled and contains intrinsic topological order, and
show that its gapless edge state is robust against symmetry-breaking perturbations. These properties of the Gutzwiller wave functions are directly confirmed numerically using the standard Monte Carlo method.

Remarkably, before projection, the above two states are both chiral at the mean-field level, but after projection the SPT state becomes non-chiral and the chiral spin liquid remains chiral.

The remaining part of this paper is organized as follows. In sections \ref{sec2} and \ref{sec3}, we briefly review the parton construction of Gutzwiller projected wavefunctions, and introduce their low energy effective field theory under two different approximations. Readers who are only interested in numerical results may go directly to section \ref{sec4}, where we show that the GWF we are studying has (1) gapped short-range correlation in  the bulk, (2) zero topological entanglement entropy and unique ground state, (3) nontrivial Hall conductance, (4) symmetry protected gapless boundary. Section \ref{sec5} is devoted to a summary.

\section{mean-field Theory of Parton Construction and its Effective Field Theory}\label{sec2}

There are two approximations to calculate the low energy effective theory from the parton construction: the mean-field approach and the  Gutzwiller projection approach. They represent two different approximations. %{\color{blue} 
The mean-field approach is simple, but for some systems it captures the main physical picture given that the mean-field parameters are chosen properly. The disadvantage is that the Hilbert space has been enlarged and local quantum fluctuations are neglected. To obtain better results, one needs to go beyond the mean-field approximation and couple the partons to internal gauge fields. This problem is partially solved in the Gutzwiller approach, where the mean-field states are projected onto the original Hilbert space. %}
In this section, we will introduce the mean-field approach of the parton construction, while the Gutzwiller-projection approach will be introduced in section \ref{sec3}.

\subsection{Parton construction}\label{sec:proj}

We adopt the fermionic representation %{\color{blue} 
(see the review paper \onlinecite{PALee06RMP}, and references therein)%} 
of spin operators $\hat {S}_i^\alpha=F_i^\dag  {\mathcal S}^\alpha F_i$ with $\alpha=x,y,z$. In the case of $S=1$, $F_i=(f_{1i},f_{0i},f_{-1i})^T$, $\mathcal S^\alpha$ are $3\times3$ matrices, and the three spin bases are represented as $|m\rangle=f_m^\dag|\mathrm{vac}\rangle$ with $m=1,0,-1$.\cite{LiuZhouNg2011_FermionMF, LiuZhouNg2011_Spin-1SL, LiuZhouTuWenNg2012, PALee12Gutz} Here a
particle number constraint $\hat N_i=f_{1i}^\dag f_{1i}+f_{0i}^\dag
f_{0i}+f_{-1i}^\dag f_{-1i}=1$ should be imposed to ensure that the Hilbert
space of fermions is the same as that of the spin. Notice that the spin operator is
invariant under the following $U(1)$ gauge transformation $F_{i}\to
F_{i}e^{i\varphi_i}$.

From the fermionic representation of $S = 1$ spin operators, 
we will consider the following pairing-free mean-field Hamiltonian on the Kagome lattice,\cite{LuLee12} 
\begin{eqnarray}\label{H_mf}
H_{\rm mf} = \sum_{ij}(t_{m,ij}e^{i \tilde a_{ij}}f_{m,i}^\dag f_{m,j} +  h.c.) 
+ \sum_i\lambda_i(\hat N_i-1),\nonumber\\
\end{eqnarray}
where the complex hopping coefficient $t_{m,ij}$ can be considered as Hubbard-Stratonovich fields in path-integral language, and the averaged value 
$\lambda_i=\bar\lambda$ is the chemical potential. Since the fermionic representation has a $U(1)$ gauge structure, the mean-field state suffers from gauge fluctuations. 
Here $(\tilde a_{ij}, \lambda_i)$ are the space and time
components of the internal $U(1)$ gauge field $\tilde a_\mu$, corresponding to  the phase fluctuations of $t_{m,ij}$ and the fluctuation of $\bar\lambda$ respectively. We integrate
out $\tilde a_\mu$ to project into the physical Hilbert space. In the mean-field approximation,  $(\tilde a_{ij}, \lambda_i)$ are not integrated out and will be fixed as $(\tilde a_{ij}, \lambda_i) =(\bar a_{ij},\bar\lambda)$.

\begin{figure}[t]
\centering
\includegraphics[width=3.5in]{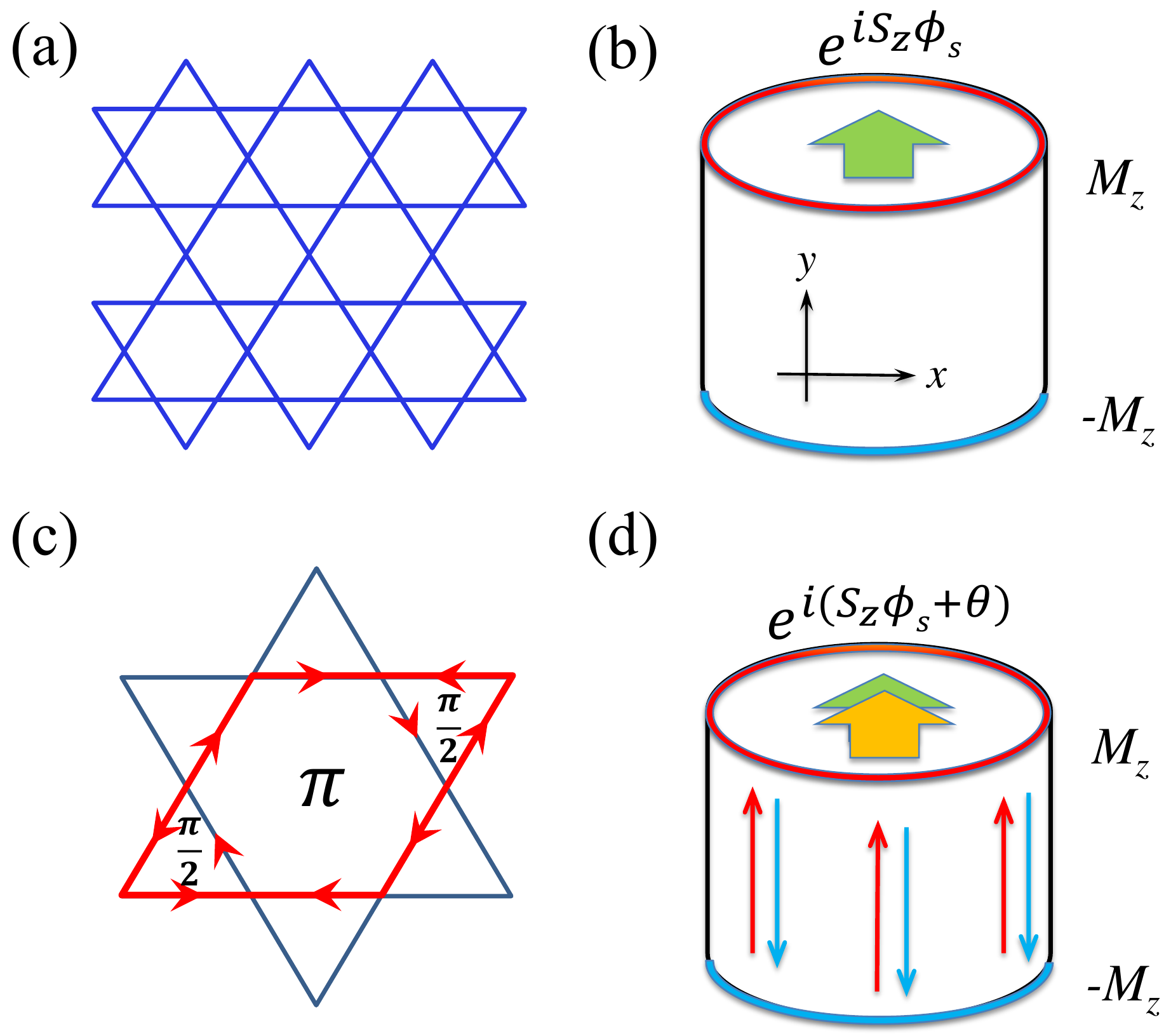}
\caption{(Color online) (a) The Kagome Lattice. (b)Laughlin's gauge invariant argument of the spin Hall conductance. The insertion of a symmetry flux through the cylinder results in symmetry charge pumping from one edge to the other [$M_z$ stands for $\sum S_{z,i}$ momentum according to the first $U(1)$ symmetry. Similarly, for the second $U(1)$ symmetry, the inserted flux should be $e^{i{S_z^2}\phi_s }$ and then $M_z$ stands for $\sum S_{z,i}^2$ momentum]. (c) The mean-field model on Kagome lattice with Chern number $\mathcal C=1$. When hopping along the arrows the fermion gain a phase $e^{i\pi/6}$; when hopping against the arrow, the fermion gain a phase $e^{-i\pi/6}$. (d) Laughlin's argument at the mean-field level. An internal gauge flux $\theta$ should be introduced such that the induced particle number flow from one edge to the other exactly cancels that caused by the symmetry flux. } \label{flux}
\end{figure}

 By tuning the phase
of $t_{m,ij}$, we can set the Chern number of each species of fermions to be
either $1$ or $-1$. For example, if we only consider nearest neighbor hopping
and set the phase to be $e^{\pm i\pi/6}$ [see Fig.~\ref{flux}(a) and \ref{flux}(c)], then the
Chern number for the lowest band is $\pm1$.  In the following discussion, we
will use the notation $|\mathcal C_1\mathcal C_0\mathcal C_{-1}\rangle$ to
denote the mean-field state, where the number $\mathcal C_m=\pm1$ stands for the Chern number of the $f_m$
species of fermion.

In the above mean-field Hamiltonian, the particle numbers of three species of
fermions are conserved respectively. This gives rise to three $U(1)$
spin symmetries. However, if the particle-number constraint is strictly satisfied, 
then the total charge (namely, the ``electric charge") degrees of freedom will be  frozen.  
As a consequence, there are two independent $U(1)$ symmetries for the spin model, 
one generated by  $\sum_i S_{z,i}$ and another by $\sum_i (S_{z,i})^2$.  In other words,
the symmetry group for the spin system is $U(1)\times U(1)$. To describe the spin system correctly, we should couple the fermions to 
the internal gauge field $\tilde a_\mu$. In the following we will give the low energy effective field theory based on the mean-field with fluctuating internal gauge fields.

\subsection{Chern-Simons theory and physical response}

Under hydrodynamic and mean-field approximations, we can introduce three Chern-Simons field $a_m$ to describe the current of the three species of fermions via $J_\mu^m = {1\over2\pi}\varepsilon^{\mu\nu\lambda}\partial_\nu a_{m,\lambda} $.
Then the mean-field theory can be described by the Chern-Simons Lagrangian 
\[
\mathcal L_{\rm MF}= -{i\over4\pi}\sum_m\mathcal C^{-1}_m \varepsilon^{\mu\nu\lambda} a_{m,\mu}\partial_\nu a_{m,\lambda}.
\] 
if we set $\tilde a_\mu=$const.  

After including fluctuating internal $U(1)$ gauge field $\tilde a_\mu$, we obtain the following low energy effective theory for the spin system 
\begin{eqnarray}\label{CS4}
\mathcal L&=& -{i\over4\pi}\sum_m\mathcal C_m^{-1} \varepsilon^{\mu\nu\lambda}a_{m\mu}\partial_\nu a_{m\lambda}
+{i\over2\pi}\sum_m \varepsilon^{\mu\nu\lambda}\tilde a_{\mu}\partial_\nu a_{m\lambda},\nonumber\\
&=& -{i\over4\pi}\varepsilon^{\mu\nu\lambda}a_{\mu}^T K \partial_\nu a_{\lambda} 
\end{eqnarray}
where $a_\mu=\left(\begin{matrix}a_{1\mu}&a_{0\mu}&a_{-1\mu}&\tilde a_{\mu}\end{matrix}\right) ^T$ and 
\begin{eqnarray*} 
K=\left(\begin{matrix}\mathcal C_1^{-1}&0&0&1\\0&\mathcal C_0^{-1}&0&1\\0&0& \mathcal C_{-1}^{-1} &1\\ 1&1&1&0\end{matrix}\right).
\end{eqnarray*}
We only kept quadratic terms and dropped the Maxwell terms.  Since $\tilde a_\mu$ can be considered as a Lagrangian multiplier, we can integrate it first and obtain an effective mutual Chern-Simons action described by a $2\times 2$ K matrix (see Appendix \ref{sec:2*2CS}).\cite{LuLee12}

If $|$det$(K)|\neq 1$ (or the signature of $K$ is not zero, where the signature of $K$ is the number of its positive eigenvalues minus the number of negative eigenvalues), then the state of the spin-1
system represented by $|\mathcal C_1\mathcal C_0\mathcal C_{-1}\rangle$ will
carry a non-trivial topological order.  If $|$det$(K)| = 1$ (or the signature of
$K$ is zero), then the corresponding spin-1 state will have a trivial topological order.
But such a state may have a non-trivial SPT order.

To detect the SPT order, we couple the system with a probe fields $A^s_\mu$ (according to some symmetry) via
\[
\mathcal L_{\rm probe} = {i\over2\pi} \varepsilon^{\mu\nu\lambda} A^s_{\mu}   Q^T \partial_\nu a_{\lambda},
\]
where $Q = ((q^s)^T, 0)^T$ , and $q^s$ is the charge carried by the fermions according to the external probe field $A^s_\mu$. For example, for the field $A^{S_z}_\mu$ that couples to the $U(1)$ charge $\sum_iS_i^z$, $q^{S_z} = (1, 0, -1)^T$,
which gives rise to
\begin{align*}
Q_{S_z}=(1,0,-1,0)^T.
\end{align*}
For the field $A^{S_z^2}_\mu$ that couples to the $U(1)$ charge $\sum_i(S_i^z)^2$, $q^{S_z^2} = (1, 0, 1)^T$,
which gives rise to
\begin{align*}
Q_{S_z^2}=(1,0,1,0)^T.
\end{align*}

Integrating out $a_\mu$ we obtain the response theory 
\begin{eqnarray}
\mathcal L_{\rm res} ={i\over4\pi} \varepsilon^{\mu\nu\lambda} Q^TK^{-1}Q A^s_\mu\partial_\nu A^s_\lambda,
\end{eqnarray}
and three Hall conductances: 
\begin{align}
\sigma^{S_z}_H & = {1\over2\pi}Q_{S_z}^TK^{-1}Q_{S_z},
\nonumber\\
\sigma^{S_z^2}_H & = {1\over2\pi}Q_{S_z^2}^TK^{-1}Q_{S_z^2},
\nonumber\\
\sigma^{S_zS_z^2}_H & = {1\over2\pi}Q_{S_z}^TK^{-1}Q_{S_z^2}.
\end{align}
If one of the above three Hall conductances is non-zero, then the spin-1 state
has a non-trivial $U(1)\times U(1)$ SPT order.

\subsection{Response mean-field theory}

When the system couples to an external probe field $A^s_\mu$, the mean-field theory should be modified accordingly. 
To get the correct response mean-field Hamiltonian, we integrate out the matter field $a_{m\mu}$ to obtain the effective Lagrangian,
\begin{eqnarray*}\label{LAa1}
\mathcal L_{\rm eff}(A,\tilde a) =  {i\over4\pi}\sum_m \mathcal C_m \varepsilon^{\mu\nu\lambda}(\tilde a_{\mu} + q_mA^s_\mu) \partial_\nu(\tilde a_{\lambda}+ q_mA^s_\lambda).%\nonumber\\
\end{eqnarray*}
The external field $A^s_\mu$ will induce a background internal gauge field $\bar a_\mu$ ----- the saddle-point value of the $\tilde a$ field which can be obtained from $\tilde J_\mu= {\delta \mathcal L_{\rm eff}(A^s,\tilde a)\over\delta \tilde a_\mu}=0$ \footnote{$\tilde J_0=0$ is nothing but the number constraint, which induces $\tilde J_i=0$. } in a proper gauge choice, 

\begin{eqnarray}\label{abar}
\bar a_\mu=-{\sum_m\mathcal C_m q_m \over \sum_m\mathcal C_m}A^s_\mu.
\end{eqnarray}
Rewriting $\tilde a_\mu=\bar a_\mu +\delta \tilde a_\mu$, 
we have
\begin{eqnarray*}
\mathcal L_{\rm eff}(A,\delta\tilde a) =  {i\over4\pi}\sum_m  \varepsilon^{\mu\nu\lambda} \mathcal C_m \left[\tilde q_m^2A_\mu^s \partial_\nu A_\lambda^s +\delta\tilde a_\mu\partial_\nu\delta\tilde a_\lambda \right],
\end{eqnarray*}
where $\tilde q_m = q_m(1-{\sum_nq_n\mathcal C_n \over q_m\sum_n\mathcal C_n})$ is the screened charge. Integrating out $\delta \tilde a_\mu$ we obtain the response Lagrangian and the spin Hall conductance is given by $\sigma^s_H={1\over2\pi}\sum_m\mathcal C_m\tilde q_m^2$.

Notice that the saddle point value $\bar a_\mu$ enters the mean-field theory, and thus the response mean-field Hamiltonian with probing field $A^s$ is given as
\begin{eqnarray}\label{ResMF}
H_\text{mf}(A^s,\bar a)&=&\sum_{m,ij}(t_{m,ij}e^{i\bar a_{ij} + iq_mA^s_{ij}}f_{m,i}^\dag f_{m,j}+\text{h.c.})\nonumber\\
&&+ \sum_i\bar a_0(N_i-1),
\end{eqnarray}
where $\bar a_\mu$ is a function of $A^s_\mu$ as given in (\ref{abar}).  Physical quantities of the spin system can be measured from the Gutzwiller projected ground state of the above mean-field Hamiltonian.

\section{Gutzwiller Construction and Effective Field Theory}\label{sec3}

\subsection{Construction of Gutzwiller wave functions}\label{sec:wavefunctions}

From the fermionic parton representation of $S = 1$ spin operators, one can construct trial spin wavefunctions for interacting spin-1 systems via Gutzwiller projection,\cite{Gros88}
\begin{eqnarray*}
|\psi\rangle_{\rm spin}=P_G|{\rm MF}\rangle,
\end{eqnarray*}
where $|{\rm MF}\rangle$ is the ground state of the mean-field Hamiltonian 
(\ref{H_mf})
and the Gutzwiller projection operator $P_G$ means only keeping the components of the mean-field state that satisfy the particle number constraint $\hat N_i=1$. Since the mean-field state suffers from gauge fluctuations, Gutzwiller projection is a simple way to partially integrate out the gauge fluctuations to obtain trial spin wave functions. For example, in 1D Gutzwiller projected $SO(3)$ symmetric $p$-wave weak pairing states belong to a nontrivial SPT phase---the Haldane phase. \cite{LiuZhouTuWenNg2012}

The above GWFs have two $U(1)$ spin symmetries, one generated by  $\sum_i S_{z,i}$ and another by $\sum_i (S_{z,i})^2$.  The projected states could be a topologically ordered state enriched by the $U(1)\times U(1)$ symmetry, or a SPT state protected by the $U(1)\times U(1)$ symmetry.

\subsection{Effective theory for projected states}

In section \ref{sec2}, we have obtained the effective Chern-Simons field theory for the
spin system from the mean-field theory, based on hydrodynamical approximation and by dropping higher-order terms in $a_\mu$. In this section, we will use a different approximation to calculate the effective-field theory from Gutzwiller projected states. Here we make much fewer approximations except assuming that the GWFs can approach very close to the true ground states.  We will show that the two approximations produce the same result.

Gutzwiller projection is equivalent to integrating out the temporal component of the internal gauge field, which result in $\delta(\sum_m f_m^\dag f_m-1)$. However, the spatial component of the internal gauge fluctuations are not completely ``integrated out". Thus, the gauge twisted boundary angles\cite{NTDW_85Twisted} (or the gauge fluxes through the holes of the torus) $\pmb\theta=(\theta_x, \theta_y)=%{\color{blue} 
(\oint \tilde{\pmb a}\cdot d\pmb l_x, \oint \tilde{\pmb a}\cdot d\pmb l_y)$ %}
can be seen as trial parameters of the GWF and should be ``integrated'' by hand. To this end, we should know the effective Lagrangian $L_{\rm eff}(\theta_x, \theta_y)$, which is given as
\begin{eqnarray} \label{Leff_H}
L_{\rm eff}(\pmb\theta) &=&  \langle P_G\psi_{\mathcal C}(\pmb\theta )|\partial_\tau|P_G\psi_{\mathcal C}(\pmb\theta )\rangle \nonumber
\\ && + \langle P_G\psi_{\mathcal C}(\pmb\theta )|H|P_G\psi_{\mathcal C}(\pmb\theta)\rangle,
\end{eqnarray}
where %{\color{blue} 
$\tau$ is the imaginary time and %} 
$|P_G\psi_{\mathcal C}(\pmb{\theta})\rangle$ is the projected mean-field state with Chern numbers $\mathcal C=(\mathcal C_1, \mathcal C_0, \mathcal C_{-1})$ and gauge twisted-boundary angles $\pmb\theta$. 

The dynamical term $\langle P_G\psi_{\mathcal C}(\pmb\theta )|H|P_G\psi_{\mathcal C}(\pmb\theta)\rangle$ is expected to be small and will be dropped in the following discussion. The consequence of the dynamic term will be discussed in section \ref{sec:TrivialTO}. The topological term $\langle P_G\psi_{\mathcal C}(\pmb\theta )|\partial_\tau|P_G\psi_{\mathcal C}(\pmb\theta )\rangle$ is the Berry phase of Gutzwiller projected states,
\begin{eqnarray*}
e^{i\oint\pmb{\mathcal A}(\pmb\theta)\cdot d\pmb\theta} &=&\exp\{\oint \langle P_G\psi_{\mathcal C}(\pmb\theta )|\partial_\tau|P_G\psi_{\mathcal C}(\pmb\theta )\rangle d\tau\}\\
&\approx& \prod\langle P_G\psi_{\mathcal C}(\pmb\theta )|P_G\psi_{\mathcal C}(\pmb\theta +\delta\pmb\theta)\rangle.
\end{eqnarray*}
The Berry %{\color{blue} 
connection $\pmb{\mathcal A(\theta)}=-i\ln \langle P_G\psi_{\mathcal C}(\pmb\theta )|P_G\psi_{\mathcal C}(\pmb\theta +\delta\pmb\theta)\rangle$ (not to be confused with the symmetry connection $\pmb A^s$) can be obtained from %} 
 the wave function overlap (see Appendix \ref{app:overlap}). Then we can calculate the Berry curvature $\mathcal F(\pmb\theta)=\partial_{\theta_x}\mathcal A_y-\partial_{\theta_y}\mathcal A_x$ and the Chern number on the torus formed by the gauge twisted boundary angles,
\begin{eqnarray}\label{k}
k={1\over2\pi}\oint_Bd\pmb\theta\cdot\pmb{\mathcal A}(\pmb\theta)={1\over2\pi}\iint_{\rm torus}d\theta_xd\theta_y\mathcal F(\pmb\theta),
\end{eqnarray}
where $B$ is a big loop that encloses the total area of the torus. It turns out that the Berry curvature is uniform on the $(\theta_x, \theta_y)$ torus. If we treat $(\theta_x,\theta_y)$ as the coordinates of a single particle on a torus, then the Berry curvature is the magnetic field that couples to the particle, and $L_{\rm eff}(\pmb\theta)$ can be written as
\begin{eqnarray}\label{Leff}
L_{\rm eff}(\pmb\theta)=i{k\over2\pi}\dot\theta_x\theta_y,
\end{eqnarray}
where ${2\pi k}$ is the strength of the ``magnetic field" %{\color{blue} 
and the dot means $\partial_\tau$. %}
From above Lagrangian, it can be shown (see Appendix \ref{app:GSD}) that the ground state degeneracy of the system is equal to $k$.

The Hall conductance can be measured by coupling the system to a symmetry flux, or symmetry twisted angles $\pmb\phi^s= (\phi^s_x, \phi^s_y)= %{\color{blue} 
(\oint \pmb { A^s}\cdot d\pmb l_x, \oint \pmb {A^s}\cdot d\pmb l_y)$.%}
 Now the GWF depends on both $\pmb\theta$ and $\pmb\phi^s$. The effective Lagrangian is given by $L_{\rm eff}(\pmb\theta,\pmb\phi^s) \approx  \langle P_G\psi_{\mathcal C}(\pmb\theta, \pmb\phi^s)|\partial_\tau|P_G\psi_{\mathcal C}(\pmb\theta, \pmb\phi^s )\rangle$. Similar to the previous discussion, the effective Lagrangian can also be written as
\[
L_{\rm eff}(\pmb\theta,\pmb\phi^s) = {i\over2\pi}\sum_m \mathcal C_m ( \dot\theta_x + q^m \dot\phi^s_x)( \theta_y + q^m \phi^s_y).
\]
The angles $\pmb\theta$ are fluctuating and we should integrate it by hand. Rewritting $\pmb\theta=\bar{\pmb \theta} + \delta\pmb\theta$, where 
\begin{eqnarray}\label{thetabar}
\bar{\pmb\theta}=-{\sum_m\mathcal C_mq_m\over \sum_m\mathcal C_m}\pmb\phi^s  
\end{eqnarray}
is obtained from $ {\delta L_{\rm eff}\over\delta\theta_i}=0$,  we then have
\[
L_{\rm eff}(\pmb\theta,\pmb\phi^s)
=  {i\over2\pi} \sum_m \mathcal C_m\left( \tilde q_m^2\dot\phi^s_x  \phi^s_y +
 \dot{\delta\theta_x}  \delta\theta_y \right),
\]
where $\tilde q_m$ is defined previously. The first term in the bracket %{\color{blue} 
gives the physical response $\sigma^s_H={1\over2\pi}\sum_m\mathcal C_m\tilde q_m^2$ and the second term indicates the ground state degeneracy $k=\sum_m\mathcal C_m$ (see Appendix~\ref{app:GSD}).%}

The Hall conductance can be calculated from the Chern numbers. When adiabatically varying the symmetry fluxes $\pmb\phi^s$, we obtain the Berry phase
\begin{eqnarray}
e^{i\pmb{\mathcal A}(\pmb \phi^s)\delta\pmb{\phi}^s} = \langle P_G\psi_{\mathcal C} (\pmb{\phi}^s,\pmb{\bar \theta})|P_G\psi_{\mathcal C} (\pmb{\phi}^s+\delta\pmb{\phi}^s,\pmb{\bar \theta}+\delta\bar{\pmb \theta})\rangle. \nonumber
\end{eqnarray}
Integration of the Berry curvature $\mathcal F(\pmb\phi^s)=\partial_{\pmb\phi^s_x}\mathcal A_y-\partial_{\pmb\phi^s_y}\mathcal A_x $ on the $(\phi^s_x, \phi^s_y)$ torus gives the Hall conductance
\begin{eqnarray}\label{SH}
2\pi\sigma_{\rm H}=\oint_B d\pmb\phi^s\cdot \mathcal A(\pmb\phi^s)=\iint_{\rm torus} d\phi^{s}_xd\phi^{s}_y\mathcal F(\pmb\phi^s).
\end{eqnarray}
%where $B$ is a big loop that encloses the total area of the torus. 

The internal back ground gauge flux $\bar\theta$ in the above discussion [or $\bar a_\mu$ in (\ref{abar})] 
is very important. Without $\bar\theta$ (or $\bar a_\mu$), GWF will give incorrect responses. To see why $\bar\theta$ (or $\bar a_\mu$) is important, we consider the electromagnetic response as an example. It is known that a spin system is a Mott insulator having no charge response. However, if we barely couple the electromagnetic field $A^c_\mu$) to the fermions, then after the Gutzwiller projection the GWF still has dependence on $\pmb\phi^c$ (or $A^c_\mu$), and the Chern-number for the GWF on the twisted-boundary-angle torus formed by $\pmb\phi^c$ is nonzero. This seems to indicate that the system still have electromagnetic quantum Hall effect. This is obviously wrong. To obtain the correct response, we need to couple both $A^c_\mu$ and $\bar a_\mu$ to the fermions. Since $q^c=(1,1,1)^T$, from (\ref{thetabar}), $\bar{\pmb \theta} = -\pmb\phi^c$ (or $\bar a_\mu=-A^c_\mu$), so the mean-field state and the projected state are independent on $\pmb\phi^c$ (or $A^c_\mu$), which is consistent with the fact that the system is an insulator.

\section{Numerical results}\label{sec4} 

\begin{figure}[t]
\centering
\includegraphics[width=3.4in]{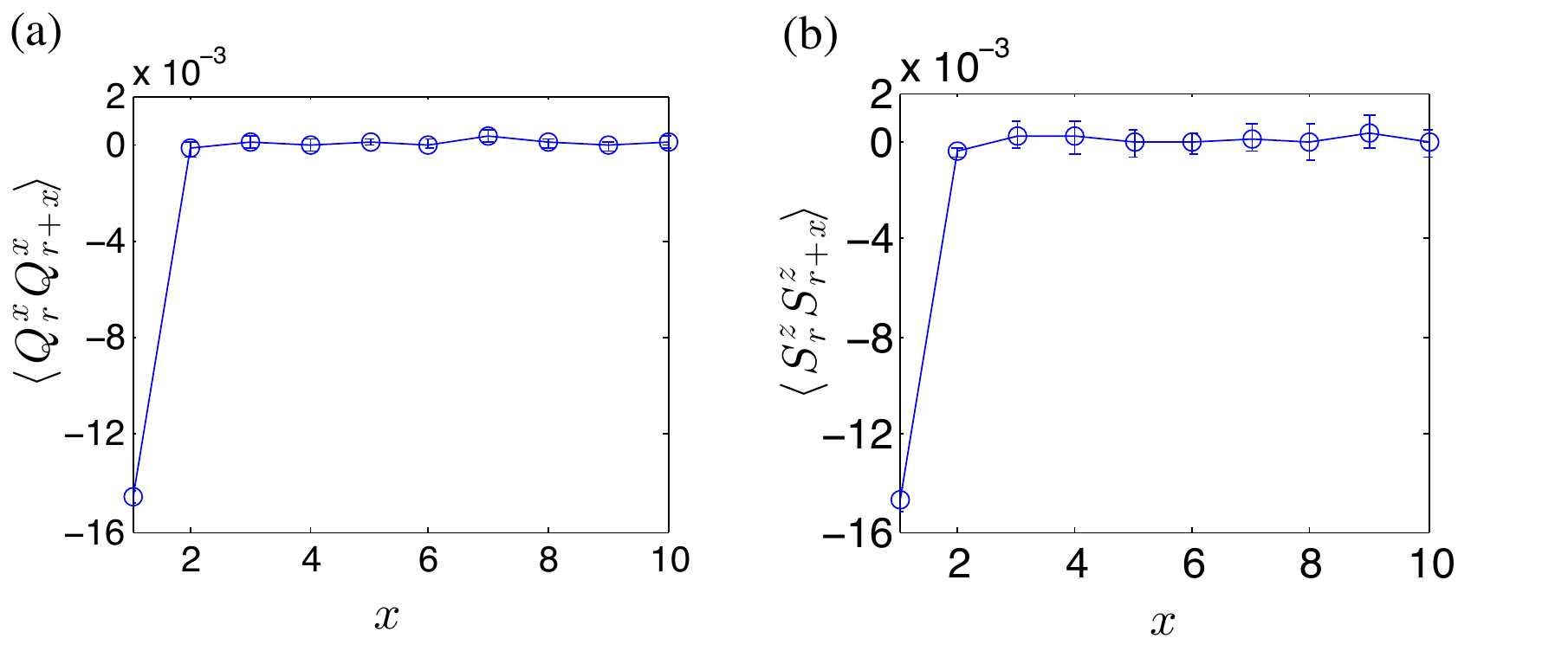}
\caption{(Color online) The correlation length on the bulk is extremely short, indicating the bulk is gapped and has no symmetry breaking. } \label{fig:bulk}
\end{figure}

In this section, we present our numerical results. We will focus on the physical properties of the state $P_G|1-11\rangle$, from which we can judge whether or not it is a SPT state.  As a comparison, the chiral spin liquid (CSL) state $P_G|111\rangle$, which carries intrinsic topological order, is also studied.

\subsection{Short range correlation in the bulk}
We first check that the bulk is gapped without symmetry breaking. To this end, we calculate the spin-spin correlation $\langle S^z_rS^z_{r+x}\rangle$ and quadrupole-quadrupole correlation$\langle Q^x_rQ^x_{r+x}\rangle$, where $Q^x=S_x^2-S_y^2$. As shown in Fig.~\ref{fig:bulk}, the correlations are weak and extremely short-ranged (about 2 lattice-constants). This indicates that the bulk has a finite excitation gap and no symmetry breaking (otherwise the correlation will be long-ranged).

\subsection{Trivial Topological Order}\label{sec:TrivialTO}

   \begin{figure}[t]
   \centering
   \subfigure[ State on a torus.]{\includegraphics[width=0.2\textwidth]{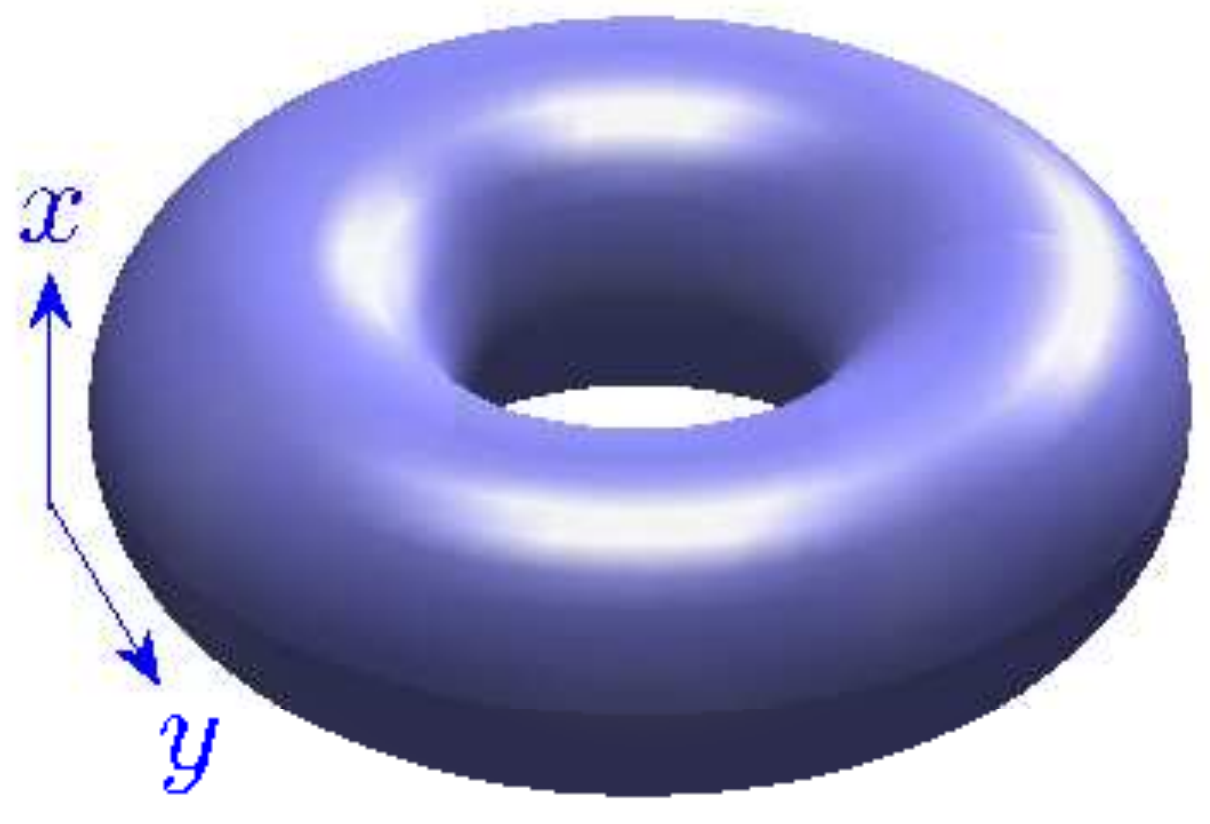}
   \label{fig_TEEa} }
   \subfigure[ Cut the torus.]{\includegraphics[width=0.2\textwidth]{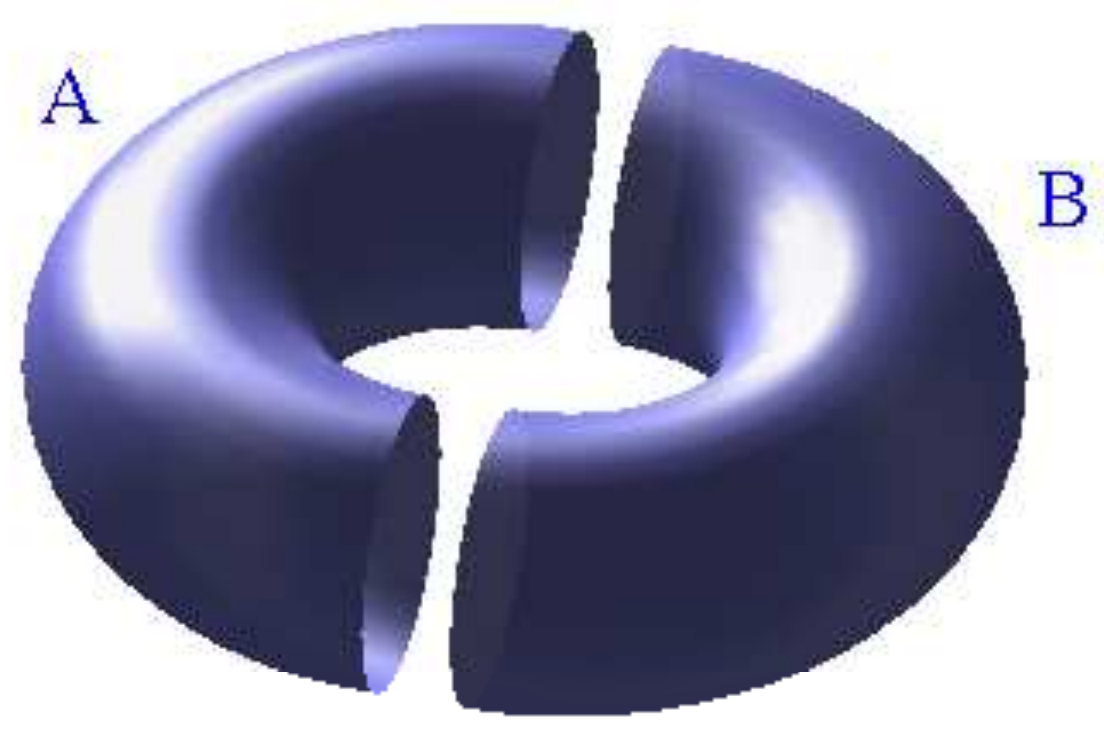}
   \label{fig_TEEb} }
   \vspace{0.1in}
   \subfigure[ Topological entanglement entropy for $P_G|1-11\rangle$. ]{\includegraphics[width=3.6in]{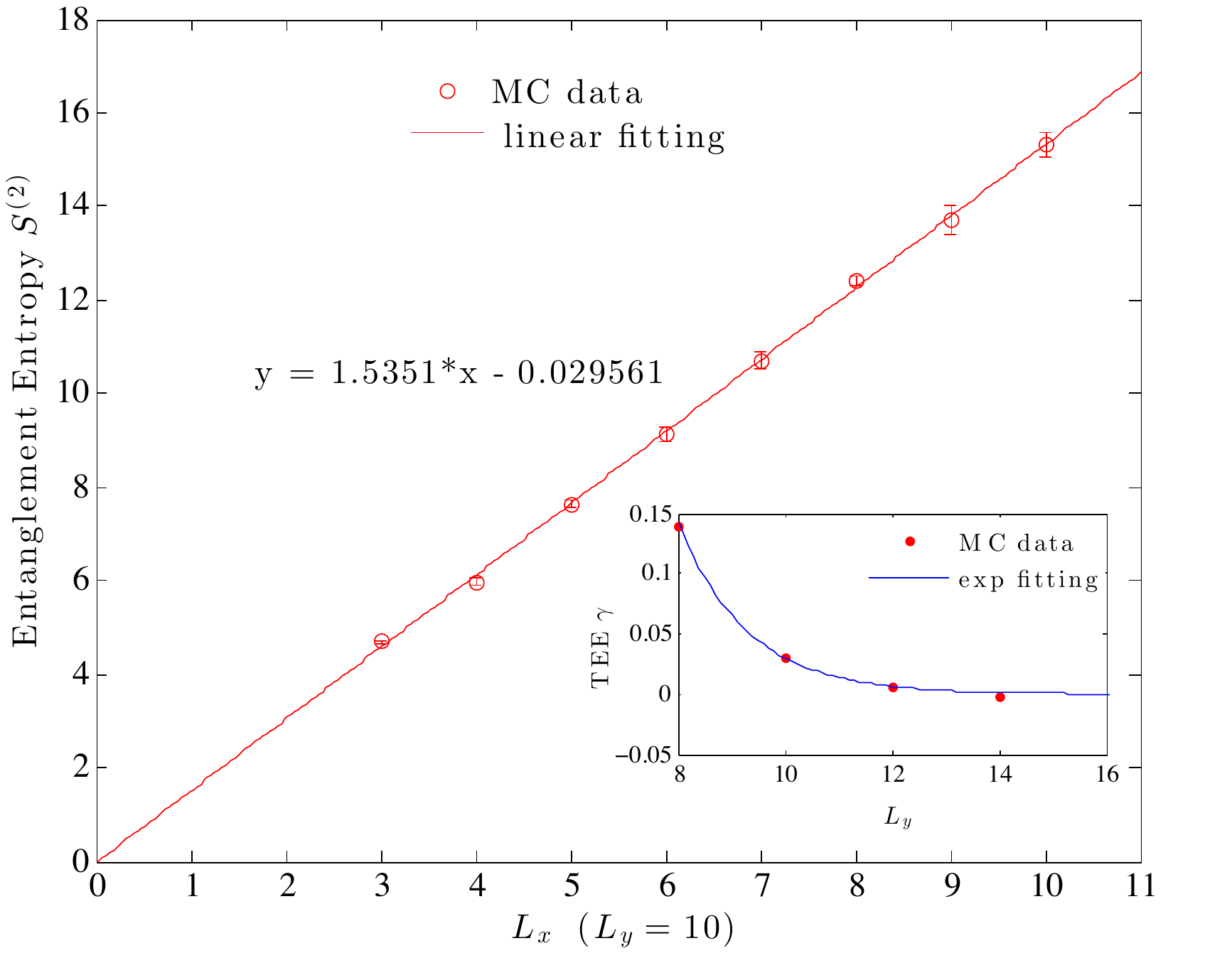}
   \label{fig_TEE} }
   \caption{Calculation of topological entanglement entropy (TEE) of state $P_G|1-11\rangle$ on a torus. (a)The geometry of the torus. (b) The torus is cut into the `system' A and the ``environment" B. (c) The second Renyi entanglement entropy $S^{(2)}=-{\rm Tr} \rho_A^2$ is plotted vs the circumference $L_x$. The intercept gives the  TEE $\gamma$. The inset shows that $\gamma$ exponentially decays to 0 with increasing ``length"  $L_y$. }\label{TEE}
\end{figure}

Here we check if the state $P_G|1-11\rangle$ has topological order by calculating its  topological entanglement entropy (TEE) and ground state degeneracy.

Using the Monte Carlo method, we can obtain the TEE from the second Renyi entropy $S^{(2)} = -{\rm Tr}\rho_A^2$, \cite{Hastings_TEE, Cirac_CFT, Zhang2011PRL,Zhang2011PRB, PeiLi2013} where $\rho_A$ is the reduced density matrix of a subsystem $A$. For topologically ordered states, the entanglement entropy have an universal correction to the area law, 
\[
S^{(2)}=\alpha \mathbb A-\gamma,
\] 
where $\mathbb A$ is the area of the boundary of the subsystem $A$, and $\gamma$ is called the topological entanglement entropy. If  $P_G|1-11\rangle$ is a SPT state (which is short range entangled), its TEE $\gamma$ should be zero. 

This is checked numerically. We consider a torus and cut it along the $x$ direction to divide it into two pieces, where each piece contains two noncontractable boundaries [see Fig.~\ref{fig_TEEb}].  Area law suggests that the second Renyi entanglement entropy is proportional to the circumference of the cut ($L_x$). In Fig.~\ref{fig_TEE}, we fix $L_y=10$ and plot the entropy with $L_x$. The TEE is given by the intersect, which is very close to 0. The inset shows that  the dependence of the TEE $\gamma$ on $L_y$. The result is that $\gamma$ exponentially decays to 0 with increasing $L_y$. The vanishing TEE implies that the state $P_G|1-11\rangle$ is indeed topologically trivial.

The trivial topological order carried by $P_G|1-11\rangle$ can also be reflected by its non-degeneracy on torus. The ground state degeneracy $k$ can be obtained through (\ref{k}). Our numerical result shows the Chern number of $P_G|1-11\rangle$ is 1, while the Chern number for the CSL state $P_G|111\rangle$ is 3, in agreement with theoretical prediction $k=\sum_m\mathcal C_m$.

\begin{figure}[t]
\centering
\includegraphics[width=3.6in]{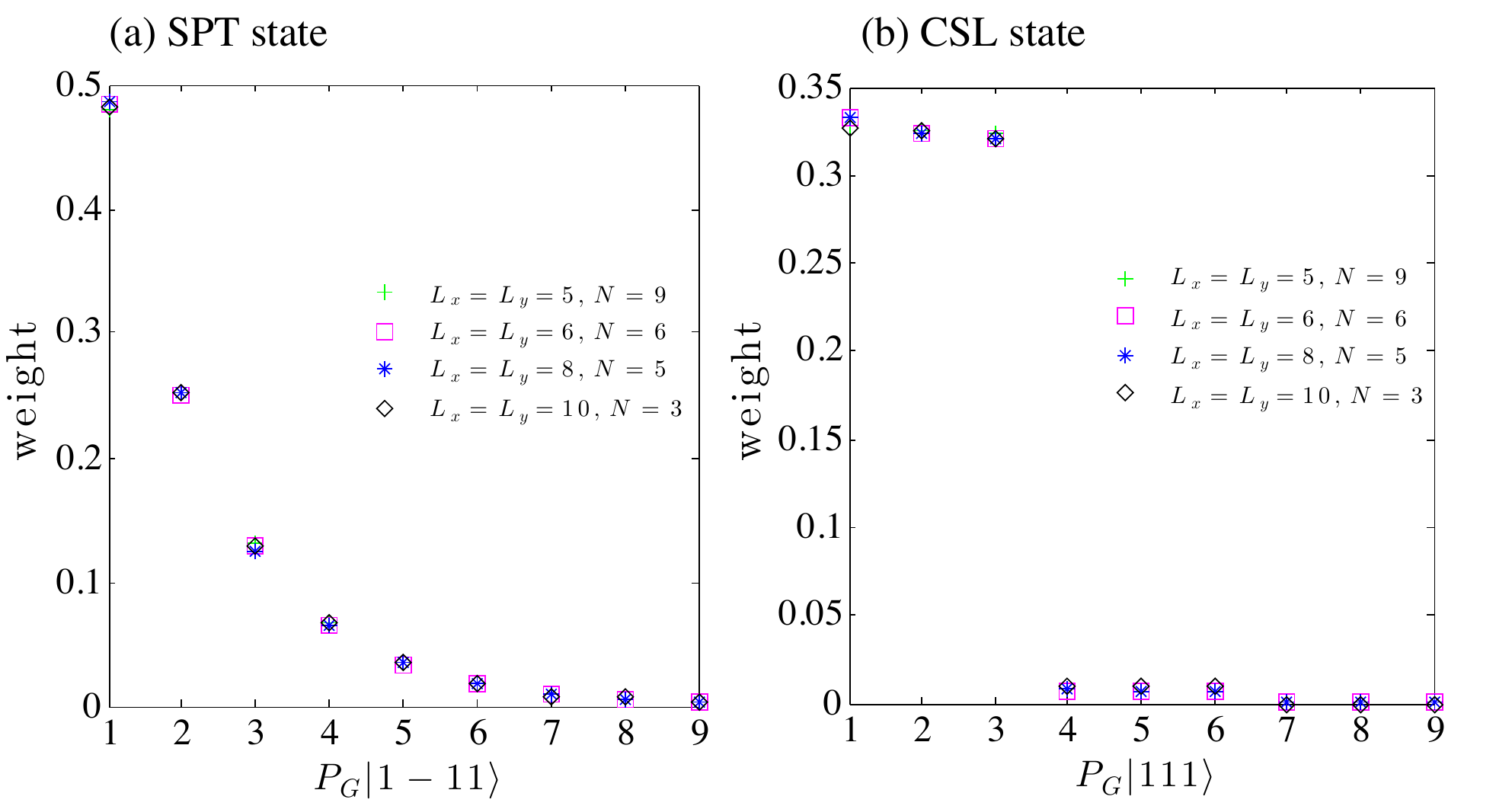}
\caption{(Color online) The largest nine (normalized) eigenvalues of the density matrix $\rho$ are shown, which almost exhaust the total weight 1. (a) Data for $P_G|1-11\rangle$; (b) data for $P_G|111\rangle$. The results are almost independent on the system size $L_x, L_y$ (the number of sites is equal to $L_x\times L_y\times3$) and the number of grids $N\times N$ by which the torus is discretized. } \label{fig_overlap}
\end{figure}

To verify that the ground state degeneracy is indeed equal to $k$, we calculate the density matrix of projected states with different twisted-boundary angles,
\begin{eqnarray}
\rho(\pmb\theta,\pmb\theta')=\langle P_G\psi_{\mathcal C}(\pmb\theta)|P_G\psi_{\mathcal C}(\pmb\theta') \rangle.
\end{eqnarray}
The eigenstates of the above density matrix are the orthogonal bases of the Hilbert space spanned by the projected states.  In numerical calculation, the torus formed by $\theta_x\in[0,2\pi)$ and $\theta_y\in[0,2\pi)$ is discretized into $N\times N$ grids. The eigenvalues of $\rho$ are proportional to the weights of the corresponding eigen-states in the GWF space. We can normalize the total weight to 1. Our data in Fig.~\ref{fig_overlap} show that the total weight is dominated by the first few states, and this result is independent on the system size and the number of grids on the $(\theta_x,\theta_y)$ torus.

If a dynamic term ${1\over g^2}(\dot\theta_x^2 + \dot\theta_y^2)$(where $g$ is a non-universal coupling constant determining the internal gauge ``photon'' gap) is added to Eq. (\ref{Leff}), then it describes a single particle moving on a torus in an uniform magnetic field with strength $2\pi k$.\cite{Wen89} The eigen states are Landau levels and the lowest Landau level correspond to the ground state of the spin system. When $g\to\infty$, the gap is infinitely large and only the ground states remain. Generally $g$ is finite and excited states occur in the GWF space with a weight $\propto e^{-\beta \varepsilon_i}$, where $\beta$ is a constant and $\varepsilon_i$ is the energy of the $i$th excited state ({\it i.e.} the $i$th Landau level).  This is the reason why there are some small weight eigenvalues appearing in  Fig.~\ref{fig_overlap}. Furthermore, the degeneracy of eigenvalues of $\rho$ reflects the degeneracy of the Landau levels, namely, the degeneracy of eigen states of the spin system on a torus. From Fig.~\ref{fig_overlap}(b), we can learn that all the eigenvalues of $\rho$ for $P_G|111\rangle$ are three-fold degenerate (within tolerable error), so the ground state is three-fold degenerate. However, for the state $P_G|1-11\rangle$, all the eigenvalues of $\rho$ are non-degenerate, indicating that the ground state is unique.

\subsection{Even-Quantized Hall conductance} 

\begin{figure}[t]
\centering
\includegraphics[width=3.4in]{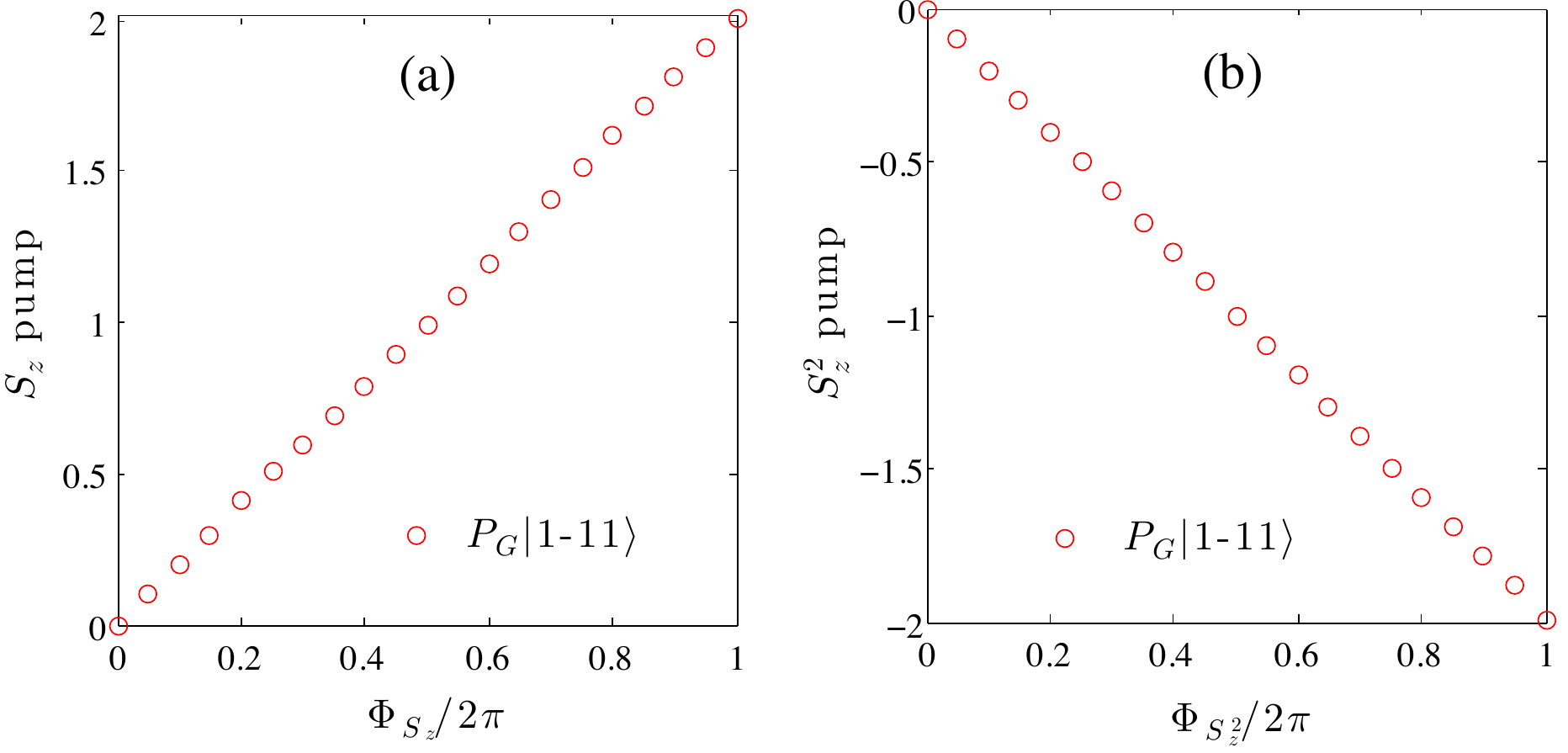}
\caption{(Color online) Symmetry charge pumping caused by inserting symmetry fluxes. The Hall conductance is equal to the charge pump by a flux quanta. (a)For the first $U(1)$ symmetry, the Hall conductance is equal to ${1\over 2\pi}2$. (b)  For the second $U(1)$ symmetry, the Hall conductance is equal to $-{1\over 2\pi}2$.} \label{fig2}
\end{figure}

We adopt Laughlin's gauge invariant argument on a cylinder to measure the Hall conductances. To this end, we adiabatically insert a $U(1)$ symmetry flux quanta $\phi^s$ into the cylinder and detect the $U(1)$ symmetry charge pumped from the bottom boundary to the top boundary. Since there are two $U(1)$ symmetries, we measure the Hall conductance respectively. During the measurement, we used the response mean-field Hamiltonian (\ref{ResMF}) to obtain the GWFs. Our numerical results of $\sigma_H^{S_z}$ and $\sigma_H^{S_z^2}$ are shown in Fig.~\ref{fig2} and the crossed Hall conductance $\sigma_H^{S_zS_z^2}$ is zero. All of the Hall conductances are even integers, consistent with  Chern-Simons theory predictions.

The spin Hall conductance can also be calculated by measuring the Chern number [see Eq. (\ref{SH})] of the projected states in the torus formed by the $U(1)$ symmetry twisted boundary angles. Our numerical results confirm the spin Hall conductance shown in Fig.~\ref{fig2}.

\begin{figure}[t]
\centering
\includegraphics[width=3.4in]{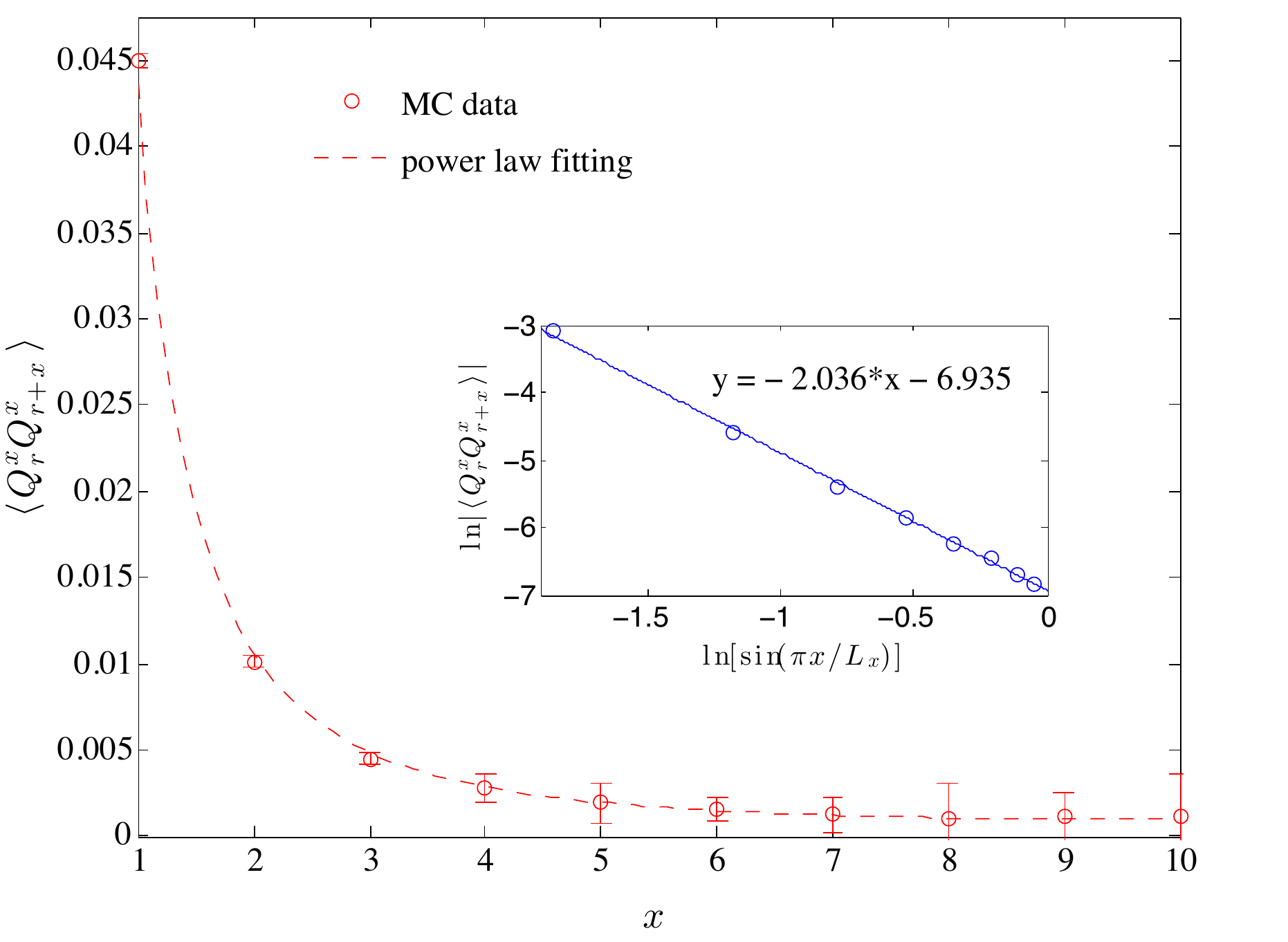}
\caption{(Color online) Power-law decaying correlation function on the boundary [the upper boundary of Figa.~{\ref{flux}}(a)-\ref{flux}(c)] shows that the edge states are gapless. Inset: Log-log fitting. The horizontal axis is set as $\ln(\sin{\pi x\over L_x})$ because of the finite-size effect.} \label{fig3}
\end{figure}

As mentioned, the spin-spin correlation function in the bulk is short ranged and boring. But the boundary is nontrivial. The nonzero Hall conductance indicates that the boundary should be gapless and the correlation function should be power law decaying. We would like to directly confirm the power law behavior for the boundary states.  We calculate the correlation function $\langle Q^{x}(r)Q^{x}(r+x)\rangle$ (where $Q^{x}=S_x^2-S_y^2$) on the boundary (along the $x$ direction) of a cylinder of 300 sites. The cylinder has $L_x\times L_y=20\times5=100$ unit cells and is periodic in the $x$ direction and open in the $y$ direction (see Fig.~\ref{flux}b). The result shows perfect power (see Fig.~\ref{fig3}), 
\[
\langle Q^{x}(r)Q^{x}(r+x)\rangle\sim x^{-2.036}
\]
and the decaying power $-2.036$ agrees well with conformal field theory prediction $-2$ (see Appendix \ref{sec:2*2CS}). It should be noted that the correlation function is very small even on the boundary. This may be due to the extremely short correlation length on the bulk.

To completely confirm that $P_G|1-11\rangle$ is a SPT state, we finally need to show that its boundary state is non-chiral, namely, the gapless boundary excitations can be gapped out by symmetry breaking perturbations. Before projection, the mean-field state $|1-11\rangle$ is obviously chiral and its boundary cannot be gapped out by small local perturbations. To show that the projected state $P_G|1-11\rangle$ is non-chiral, we calculate the boundary correlation function after adding some symmetry-breaking perturbation.

\subsection{Symmetry protected Gapless boundary states} \label{sec:gapless}

The $U(1)$ symmetry breaking perturbation that we consider is the following fermion pairing term
\begin{eqnarray}\label{Pairing}
H'_{\rm mf}= \Delta_{ij}^1 c_{1i}^\dag c_{1j}^\dag + \Delta_{ij}^2c_{-1i}^\dag c_{0j}^\dag +h.c.
\end{eqnarray}
The spin interaction that support sthis perturbation might be 
\[
H'= -(c_{1i}^\dag c_{1j}^\dag c_{-1j} c_{0i} +h.c.) = - (P^x_iQ^x_j-P^y_iQ^y_j),
\] 
where $P^x={1\over\sqrt2}(S_xS_z+S_zS_x+S_xS_y)=\left(\begin{matrix}0&1&0\\1&0&0\\0&0&0\end{matrix}\right)$, $P^y={1\over\sqrt2}(S_yS_z+S_zS_y+S_y)=\left(\begin{matrix}0&-i&0\\i&0&0\\0&0&0\end{matrix}\right)$, and $Q^x=S_x^2-S_y^2=\left(\begin{matrix}0&0&1\\0&0&0\\1&0&0\end{matrix}\right)$, $Q^y=S_xS_y+S_yS_x=\left(\begin{matrix}0&0&-i\\0&0&0\\i&0&0\end{matrix}\right)$. Similar to $S^x,S^y,S^z$, the three operators $Q^x, Q^y, S^z$ also form $SU(2)$ algebra. 

\begin{figure}[t]
\centering
\includegraphics[width=3.4in]{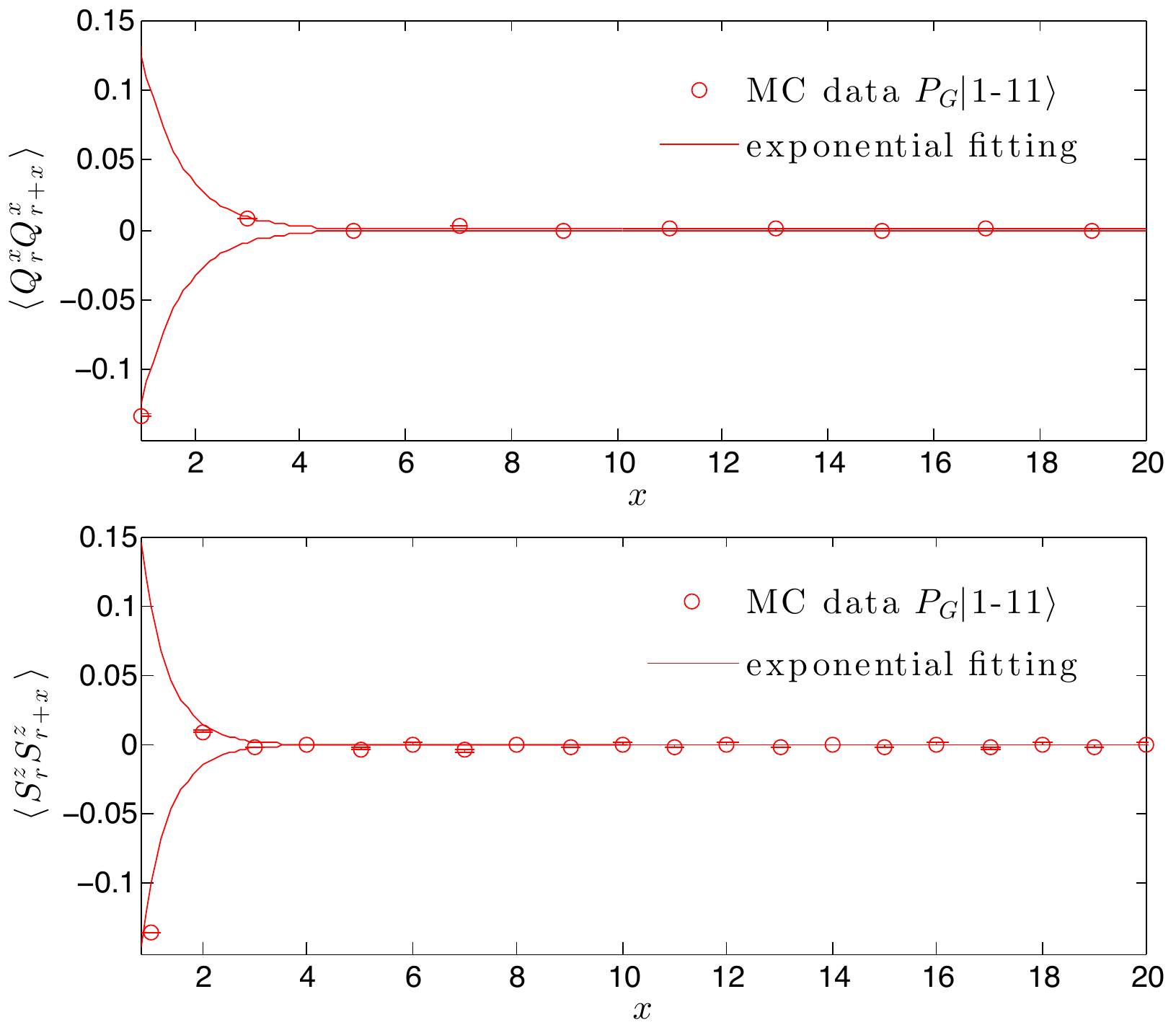}
\caption{(Color online) Boundary of SPT phase can be gapped when symmetry is explicitly broken. } \label{fig_gapSPT}
\end{figure}

Our numerical result is shown in Fig.~\ref{fig_gapSPT}, where the correlation function $\langle S_r^z S_{r+x}^z\rangle$ and  $\langle Q_r^x Q_{r+x}^x\rangle$ are both exponentially decaying as expected.

\begin{figure}[t]
 \centering
 \includegraphics[width=3.4in]{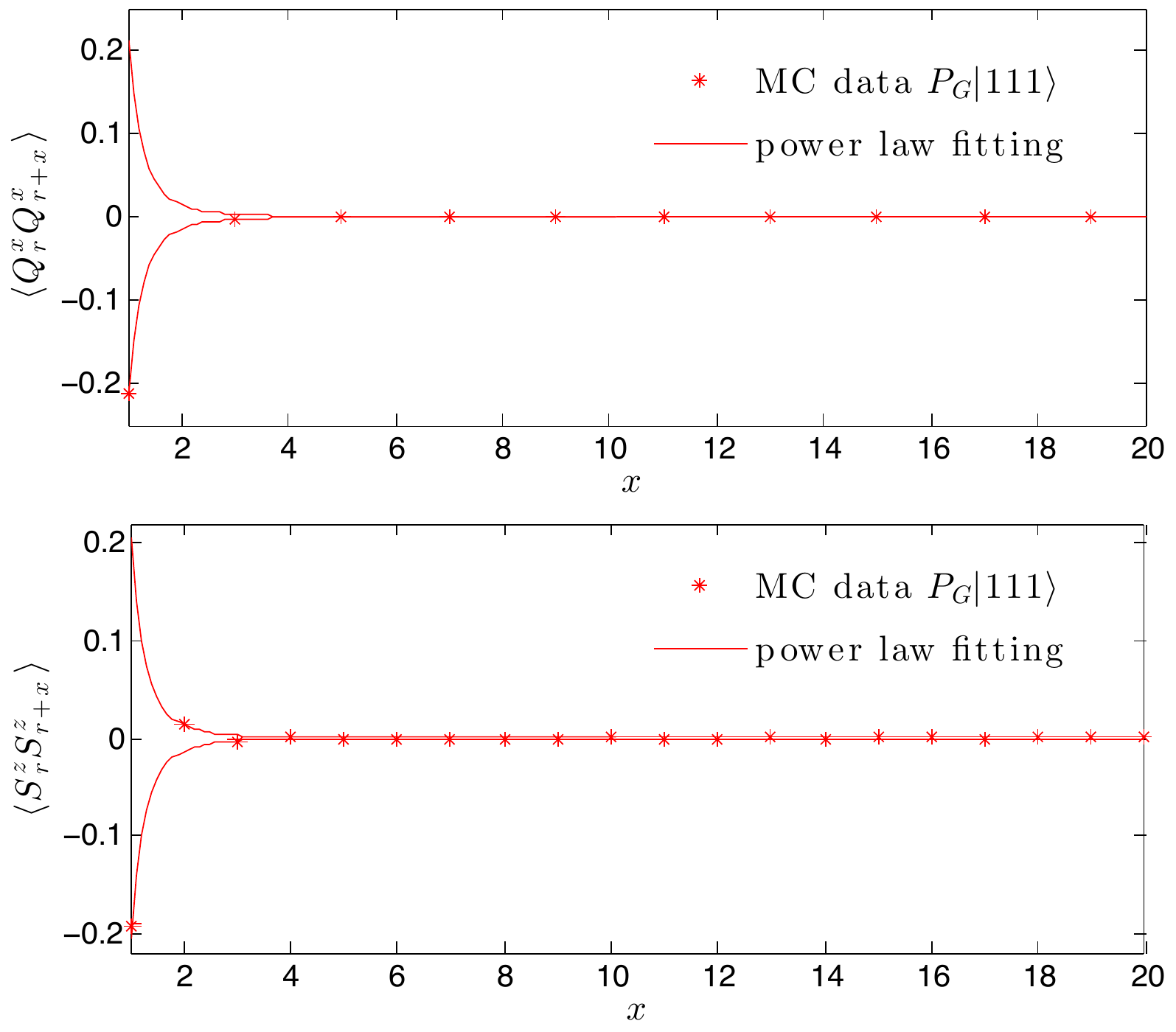}
 \caption{(Color online) Boundary of CSL phase is robust against all perturbations. } \label{fig_gaplessCSL}
 \end{figure}

We also calculate the boundary correlation function of the CSL undergoing the same perturbation.  The results in Fig.~\ref{fig_gaplessCSL} show that the boundary remains gapless under the perturbation. This comparison give strong evidence that the boundary of the state $P_G|1-11\rangle$ is non-chiral while the CSL state $P_G|111\rangle$ is chiral, as predicted by Chern-Simons theory (see Appendix \ref{sec:2*2CS}).

\section{Conclusion and discussion}\label{sec5}

In summary, using the Monte Carlo method we studied the physical properties of Gutzwiller-projected wave functions. We especially studied the state $P_G|1-11\rangle$ (where $1,-1,1$ are the mean-field Chern numbers of the fermions $f_1, f_0, f_{-1}$, respectively), including its spin Hall conductance, correlation function of the gapless edge states,  ground-state degeneracy and topological entanglement entropy, and nonrobustness of the gapless edge states. All these evidences show that $P_G|1-11\rangle$ is a $U(1)\times U(1)$ symmetry protected topological state. Our work may shed some light on simple lattice models and experimental realization of SPT phases.

The spin Hall conductance is calculated by measuring the spin pump in the Gutzwiller wave function caused by inserting symmetry flux through the cylinder to the mean-field Hamiltonian. We find that the internal gauge field plays an important role since external symmetry flux will induce a nonzero background internal gauge flux (see also Ref.~\onlinecite{MeiWen14}). 
Our observation indicates that in general the internal gauge field cannot be ignored in studying the physical response of Gutzwiller projected wave functions. 

We also compared the SPT state $P_G|1-11\rangle$ with the topologically ordered chiral spin liquid state $P_G|111\rangle$ whose gapless boundary excitations are robust against all local perturbations. Our data imply that the boundary of the SPT state is non-chiral while the boundary of the chiral spin liquid is chiral. Noticing that at the mean-field level both $|1-11\rangle$ and $|111\rangle$ are chiral, it is remarkable that after Gutzwiller projection (or due to strong interactions) the former becomes non-chiral. This indicates that the physical properties of some mean-field states might be dramatically changed after Gutzwiller projection.

Our Gutzwiller approach can be applied to study SPT states protected by other symmetry groups, such as $SU(2)$ or $SO(3)$ symmetry, and so on. It can be also used to study symmetry enriched topological phases, where symmetry interplays with topological order resulting in an enriched phase diagram. 

%{\color{blue} 
Finally, we give some remarks about the Hamiltonians of the Gutzwiller-projected states that we constructed above. In principle, for each Gutzwiller wavefunction one can always find a parent Hamiltonian of which the Gutzwiller wavefunction is the ground state. However, that Hamiltonian is generally very complicated and is difficult to identify. Nevertheless, approximate Hamiltonians can be constructed. For instance, in Ref.~\onlinecite{LuLee12} a spin Hamiltonian containing three-body interactions was proposed via perturbation to onsite Hubbard interactions. On the other hand, the reduced density matrix method introduced in Ref.~\onlinecite{MeiWen14} also provides some hint of possible interactions that may stabilize the Gutzwiller wavefunctions.
%Remarks about the Hamiltonian...
%}

\section*{Acknowledgements}

We thank Y.-M. Lu, Y. Zhang, Z.-C. Gu, H. Yao, H.-H. Tu, M. Cheng, X.-L. Qi, C. Xu and P. A. Lee for helpful discussions. This research is supported in part by Perimeter Institute for Theoretical Physics. Research at Perimeter Institute is supported by the Government of Canada through Industry Canada and by the Province of Ontario through the Ministry of Economic Development \& Innovation. Z.X.L. is thankful for the support from NSFC Grant No. 11204149 and the Tsinghua University Initiative Scientific Research Program. X.G.W. is also supported by NSF Grant No.  DMR-1005541 and NSFC Grant No. 11274192, and the John Templeton Foundation Grant  No. 39901.

\appendix

\section{Topological Entanglement Entropy}\label{app:TEE}
In Ref.~\onlinecite{Hastings_TEE, Cirac_CFT, Zhang2011PRL,Zhang2011PRB, PeiLi2013}, several tricks has been introduced to calculate Renyi entropy. The main trick is the ``sign trick" which separates the calculation of the magnitude and phase of the swap operator:
\begin{eqnarray}
e^{-S_2}=\langle \mathrm{SWAP}\rangle&=&\sum_{\alpha_1\alpha_2}\rho_{\alpha_1}\rho_{\alpha_2}{\phi_{\beta_1}\phi_{\beta_2}\over \phi_{\alpha_1}\phi_{\alpha_2}}\nonumber\\
&=&\langle \rm{SWAP}\rangle_{\rm{amp}}\langle \rm{SWAP}\rangle_{\rm{phs}},
\end{eqnarray}
where $\alpha_1, \alpha_2$ are the spin configurations of two independent systems of the same size, $\beta_1, \beta_2$ are the spin configurations after the swapping of the spins in the holes, and 
\begin{eqnarray}
\langle \rm{SWAP}\rangle_{\rm{phs}}=\sum_{\alpha_1\alpha_2}\tilde\rho_{\alpha_1,\alpha_2}e^{i\phi}
\end{eqnarray}
with  $\phi=\mathrm{Arg} (\phi^*_{\alpha_1}\phi^*_{\alpha_2}\phi_{\beta_1}\phi_{\beta_2})$ and $\tilde\rho_{\alpha_1,\alpha_2}={|\phi^*_{\alpha_1}\phi^*_{\alpha_2}\phi_{\beta_1}\phi_{\beta_2}|\over \langle\rm{SWAP}\rangle_{\rm{amp}}}$, 
\begin{eqnarray}
\langle\rm{SWAP}\rangle_{\rm{amp}}&=&\sum_{\alpha_1,\alpha_2}|\phi^*_{\alpha_1}\phi^*_{\alpha_2}\phi_{\beta_1}\phi_{\beta_2}|\nonumber\\
&=&\sum_{\alpha_1\alpha_2}\rho_{\alpha_1}\rho_{\alpha_2}\left|{\phi_{\beta_1}\phi_{\beta_2}\over \phi_{\alpha_1}\phi_{\alpha_2}}\right|
\end{eqnarray}
When calculating the phase part, since both the spin configurations before and after the swapping appear in the sampling weight, the trick of updating the inverse and determinant can be applied in the Monte Carlo steps. However, this trick can not be applied to the magnitude part since the swapped configuration may have zero weight and $\phi_{\beta_1}$, $\phi_{\beta_2}$ may not change continuously. To solve this problem and to decrease the error, here we further use the trick to separate the calculation of the magnitude into two steps, in each step, the matrix inverse and determinant updating techniques can be applied. The main idea is to introduce a weight function $f(\alpha_1,\alpha_2)$,
\begin{eqnarray}
f(\alpha_1,\alpha_2)=
\left\{
\begin{array}{rl}
1,&\ \rm {if}\ \beta_1,\ \beta_2\ \rm{are\ allowed}  \\
0,&\ \rm {if}\ \beta_1,\ \beta_2\ \rm{are\ not\ allowed}
\end{array}
\right.
\end{eqnarray}
such that
\begin{eqnarray}
\langle\rm{SWAP}\rangle_{\rm{amp}}&=&\sum_{\alpha_1\alpha_2}f(\alpha_1,\alpha_2)\rho_{\alpha_1}\rho_{\alpha_2}\left|{\phi_{\beta_1}\phi_{\beta_2}\over \phi_{\alpha_1}\phi_{\alpha_2}}\right|\nonumber\\
&=&\sum_{\alpha_1\alpha_2}\rho'(\alpha_1,\alpha_2)\left|{\phi_{\beta_1}\phi_{\beta_2}\over \phi_{\alpha_1}\phi_{\alpha_2}}\right|\langle f(\alpha_1,\alpha_2)\rangle\nonumber\\
&=&\langle\rm SWAP\rangle'_{\rm amp}\langle f(\alpha_1,\alpha_2)\rangle
\end{eqnarray}
where
$\rho'(\alpha_1,\alpha_2)={f(\alpha_1,\alpha_2)\rho_{\alpha_1}\rho_{\alpha_2}\over \langle f(\alpha_1,\alpha_2)\rangle}$, and
\[
\langle f(\alpha_1,\alpha_2)\rangle=\sum_{\alpha_1,\alpha_2}\rho_{\alpha_1}\rho_{\alpha_2}f(\alpha_1,\alpha_2).
\]
Since $f(\alpha_1,\alpha_2)$ is a simple function taking values 0 and 1, the fluctuation is reduced considerably compared to $\langle \rm{SWAP}\rangle_{\rm amp}$ itself.

\section{Overlap of wave functions }\label{app:overlap}
Suppose two normalized wave functions $|\psi_1\rangle$ and  $|\psi_2\rangle$ are given as
\begin{eqnarray*}
&&|\psi_1\rangle=\sum_{\alpha} {f_1(\alpha)\over\sqrt{\sum_\beta |f_1(\beta)|^2}} |\alpha\rangle,\\
&&|\psi_2\rangle=\sum_{\alpha} {f_2(\alpha)\over\sqrt{\sum_\beta |f_2(\beta)|^2}} |\alpha\rangle,
\end{eqnarray*}
where $\alpha$ means a spin configuration. To calculate the overlap between the two states $\langle\psi_1|\psi_2\rangle$, we introduce another normalized wave function $|\psi_0\rangle$ to generate the Monte Carlo sequence,
\begin{eqnarray*}
|\psi_0\rangle=\sum_{\alpha} {h(\alpha)\over\sqrt{\sum_\beta |h(\beta)|^2}} |\alpha\rangle = \sum_\alpha W_\alpha|\alpha\rangle,
\end{eqnarray*}
where $W_\alpha = {h(\alpha)\over\sqrt{\sum_\beta |h(\beta)|^2}}$ is the weight of $\alpha$.

Now we have
\begin{eqnarray}
\langle\psi_1|\psi_2\rangle &=& \sum_\alpha {f^*_1(\alpha)f_2(\alpha) \over\sqrt{\sum_\beta|f_1(\beta)|^2\sum_\gamma|f_2(\gamma)|^2}}\nonumber\\
&=& \sum_\alpha W_\alpha{f^*_1(\alpha)f_2(\alpha) \over h^*(\alpha) h(\alpha)}{ \sum_\sigma |h(\sigma)|^2\over\sqrt{\sum_\beta|f_1(\beta)|^2\sum_\gamma|f_2(\gamma)|^2}}\nonumber\\
&=& {1\over C} \sum_\alpha W_\alpha{f^*_1(\alpha)f_2(\alpha) \over h^*(\alpha) h(\alpha)},
\end{eqnarray}
where $C$ is a constant:
\begin{eqnarray}
C&=& \sqrt{ {\sum_\beta|f_1(\beta)|^2\over \sum_\sigma |h(\sigma)|^2}{\sum_\gamma|f_2(\gamma)|^2\over \sum_\delta |h(\delta)|^2} }\nonumber\\
&=& \sqrt{ \sum_\beta W_\beta \left|{f_1(\beta)\over h(\beta)}\right|^2 \sum_\gamma W_\gamma \left|{f_2(\gamma)\over h(\gamma)}\right|^2}
\end{eqnarray}

\section{Ground State Degeneracy and Boundary theory}\label{app:CS}

\subsection{Ground State Degeneracy}\label{app:GSD}

If we integrate out the $a_{m\mu}$ fields in the Chern-Simons action (\ref{CS4}), we obtain 
\begin{eqnarray}\label{CS_GSD}
\mathcal L_{\rm eff} (\tilde a)=  {i\over4\pi}k \varepsilon^{\mu\nu\lambda}\tilde a_{\mu}\partial_\nu \tilde a_{\lambda},
\end{eqnarray}
where $k=\sum_m\mathcal C_m$. If we further integrate out the $\tilde a_0$ field, we obtain a zero-strength condition 
\[
\partial_x \tilde a_y -\partial_y \tilde a_x=0.
\] 
So we can write $\tilde a_i=\partial_i\Lambda+\theta_i/L_i$, where $L_i$ is the size along the $i$ direction and $\theta_i$ can be interpreted as the angle of twisted boundary condition for the fermionic spinons, or the gauge flux through the $i$th hole of the torus. Substituting the above expression into (\ref{CS_GSD}), we get the effective action
\begin{eqnarray}\label{thetaxy}
L_{\rm eff} =  {i\over2\pi}k \dot\theta_{x}\theta_{y},
\end{eqnarray}
which yields $[\theta_x, {k\over2\pi} \theta_y]=i$. Define operators $T_i=e^{i\theta_i}$, then we have
\begin{eqnarray}\label{TxTy}
T_xT_y=T_yT_xe^{i{2\pi\over k}},
\end{eqnarray}
which form a Heisenberg algebra. 

Noticing $\tilde a_0$ is simply the chemical potential $\lambda_i$ in (\ref{H_mf}), integrating out $\tilde a_0$ results in exactly one fermion per site, which is equivalent to a Gutzwiller projection. Equation (\ref{thetaxy}) shows that the GWF still has some degrees of freedom, which determines the ground state degeneracy. 

The representation space of the above Heisenberg algebra (\ref{TxTy}) is at least $k$-dimensional. Since $\tilde a_\mu$ is a gauge degree of freedom for the original spin model, $T_x$ and $T_y$ will not change the spin Hamiltonian, namely, $[T_x, H]=[T_y,H]=0$. So the Hilbert space of each energy level forms a representation space of the Heisenberg algebra. In other words, all of the energy levels, including the ground state, are at least $k$-fold degenerate.  

The degeneracy of the ground states can be obtained by calculating the Chern number for the Gutzwiller projected mean-field states. At the mean-field level, $\theta_x$ and $\theta_y$ are commuting, so we can construct mean-field states with certain values of $\theta_x,\theta_y$, noted as $|\psi_{\mathcal C}({\theta_x,\theta_y})\rangle$, where $\mathcal C$ denotes $(\mathcal C_1\mathcal C_0\mathcal C_{-1})$ for short. The topological term (\ref{thetaxy}) plays its role when $\tilde a_0$ is integrated out (or, equivalently, after the Gutzwiller projection). If we interpret the topological term (\ref{thetaxy}) as the Berry phase of the Gutzwiller projected state evolving on the $(\theta_x, \theta_y)$ torus,
\begin{eqnarray}
{i\over2\pi}k \dot\theta_{x}\theta_{y}=\langle P_G\psi_{\mathcal C}(\theta_x,\theta_y)|\partial_\tau|P_G\psi_{\mathcal C}(\theta_x,\theta_y)\rangle,
\end{eqnarray}
then $k$ corresponds to the Chern number of the projected state,
\begin{eqnarray}\label{PGChern}
2\pi k=\oint{k\over2\pi}\dot\theta_{x}\theta_{y}d\tau &=& \oint_B d\pmb{\theta}\cdot\pmb{\mathcal A}\\&=&\oint_{\rm torus} d\theta_xd\theta_y \mathcal F(\pmb\theta),
\end{eqnarray}
where $\pmb{\mathcal A}=-i\langle P_G\psi_{\mathcal C}(\pmb{\theta})|\partial_{\pmb{\theta}} |P_G\psi_{\mathcal C}(\pmb{\theta})\rangle$ is the Berry connection (if $\theta_x, \theta_y$ are discretized, then we have $e^{i\pmb{\mathcal A}\cdot\delta\pmb{\theta}}=\langle P_G\psi_{\mathcal C}(\pmb{\theta})|P_G\psi_{\mathcal C}(\pmb{\theta}+\delta\pmb{\theta})\rangle$) and $\mathcal F(\pmb\theta)=\partial_{\theta_x}\mathcal A_y-\partial_{\theta_y}\mathcal A_x$ is the Berry curvature,  $B$ is the big loop enclosing the total area of the $(\theta_x, \theta_y)$ torus.

Generally, the projected state $|P_G\psi_{\mathcal C}(\theta_x,\theta_y)\rangle$ is not an eigenstate of $T_i=e^{i\theta_i}$. In stead, an eigen state $|n_i\rangle$ of $T_i|n\rangle_i=e^{i2n\pi/k}|n\rangle_i$ (here $n=0,1, ... ,k-1$) is a superposition of $|P_G\psi_{\mathcal C}(\theta_x,\theta_y)\rangle$,
\begin{eqnarray}
|n\rangle _i= \int d\theta_x d\theta_y \xi_{n_i}(\theta_x,\theta_y)|P_G\psi_{\mathcal C}(\theta_x,\theta_y)\rangle,
\end{eqnarray}
where $\xi_{n_i}(\theta_x,\theta_y)$ is a weight function in analog to a single particle wavefunction in the first Landau level. \cite{Wen89}

\subsection{Boundary theory}\label{sec:2*2CS}

In the remaining part, we will introduce an equivalent $K$-matrix description as the low energy effective theory. 
Integrating out the internal gauge field $\tilde a_\mu$ first, we obtain $\sum_m \partial_\nu a_{m\lambda}=0$, or $\sum_m a_{m\lambda}=0$ up to a constant field. Eliminating $a_{0\mu}$, we obtain the low energy effective Chern-Simons theory for the spin system,\cite{LuLee12}
\begin{eqnarray}\label{CS2}
\mathcal L&=& -{i\over4\pi} \varepsilon^{\mu\nu\lambda} \left(\begin{matrix}a_{1\mu}&a_{-1\mu}\end{matrix}\right) K \partial_\nu \left(\begin{matrix}a_{1\lambda}\\ a_{-1\lambda}\end{matrix}\right)\nonumber 
\\&&+{i\over2\pi} \varepsilon^{\mu\nu\lambda} A^s_{\mu} \left(\begin{matrix}q_1&q_{-1}\end{matrix}\right) \partial_\nu \left(\begin{matrix}a_{1\lambda}\\ a_{-1\lambda}\end{matrix}\right),
\end{eqnarray}
where $K=\left(\begin{matrix}\mathcal C_1^{-1} + \mathcal C_0^{-1}&\mathcal C_0^{-1}\\ \mathcal C_0^{-1} & \mathcal C_{-1}^{-1} + \mathcal C_0^{-1}\end{matrix}\right)$,  $A^s_\mu$ is the probing field according to some symmetry and $q=\left(\begin{matrix}q_1\\ q_{-1}\end{matrix}\right)$ is the ``charge vector'' coupling to this probe field. For the $\sum_iS^z_i$ conservation symmetry, $ q=\left(\begin{matrix}1\\ -1\end{matrix}\right)$, while for the $\sum_i(S^z_i)^2$ conservation symmetry, $q=\left(\begin{matrix}1\\ 1\end{matrix}\right)$. The Hall conductance is given by $\sigma_H={1\over2\pi}q^TK^{-1}q$.

  \begin{figure}[t]
\centering
\includegraphics[width=3.4in]{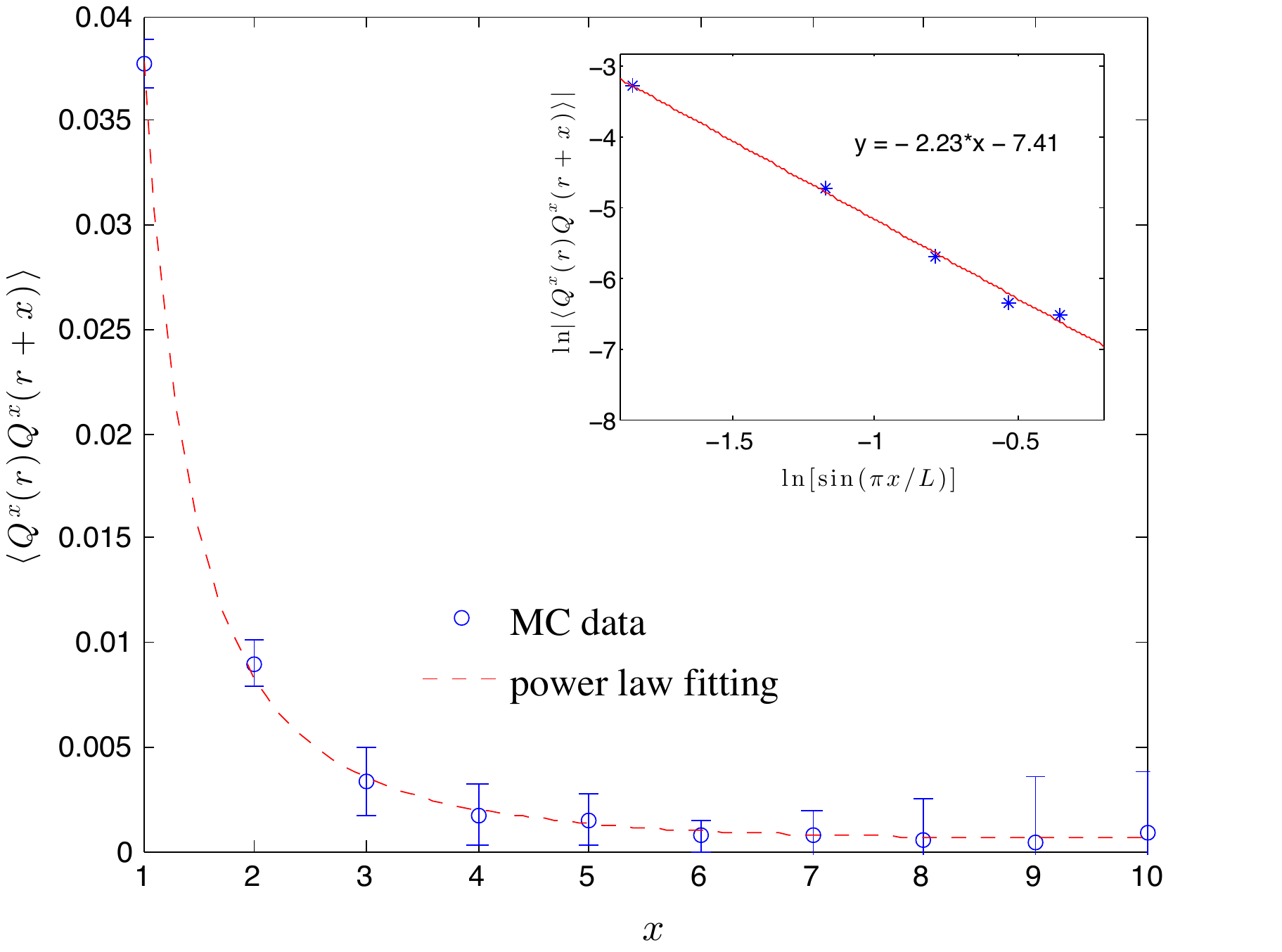}
\caption{(Color online) Correlation function of the boundary in an uniform Zeeman field $B_x=0.4$. Power law decaying correlation function on the boundary shows that the edge states are still gapless. Inset: Log-log fitting, which shows that the decaying power is approximately $-2.23$. } \label{fig:app}
\end{figure}

Since (\ref{CS2}) is not gauge invariant if the system has a boundary, we need to introduce a boundary action to recover the gauge invariance,
\begin{eqnarray}
\mathcal L_{\rm boundary} = -{i\over4\pi}K_{IJ}\partial_\tau\phi_I\partial_x\phi_J - V_{IJ}\partial_x\phi_I\partial_x\phi_J,
\end{eqnarray}
where $I,J=1,-1$, and the field $\phi_I$ only exist on the boundary and is defined such that $a_{I\mu}=\partial_\mu\phi_I$. The $\phi_I$ field satisfies the Kac-Moody algebra,
\begin{eqnarray}\label{KM}
[\partial_x\phi_I, \partial_y\phi_J ] = 2\pi i(K^{-1})_{IJ}\partial_x\delta(x-y).
\end{eqnarray}
The fermion operators can be written as $f_1\sim e^{-i\phi_1}, f_{-1}\sim e^{-i\phi_{-1}}$. The spin density operator is given as $S_z\sim\partial_x\phi_1 - \partial_x\phi_{-1}$, and 
\begin{eqnarray}\label{Q+-}
Q^{\pm}={1\over2} (Q^x\pm iQ^y) \sim e^{i\phi_{\pm1}-i\phi_{\mp1}}.
\end{eqnarray}

If $\mathcal C_1=1, \mathcal C_0=-1, \mathcal C_{-1}=1$, then the $K$ matrix is given as $K=\left(\begin{matrix}0&-1\\-1&0\end{matrix}\right)$, and from (\ref{KM}) and (\ref{Q+-}), we obtain the scaling law 
\[
\langle Q^+(r)Q^-(r+x)\rangle\sim x^{-2},
\]
which is verified by the numerical result given in section \ref{sec3}.

Furthermore, for the above $K$ matrix, since $l=\left(\begin{matrix}n\\ 0\end{matrix}\right)$ or $l=\left(\begin{matrix}0\\ n\end{matrix}\right)$ ($n$ is an integer) satisfies 
\begin{eqnarray}\label{Gapping}
l^TK^{-1}l=0,
\end{eqnarray} 
the Higgs term \cite{LuVishwanath2012} that may gap out the boundary is $\cos (n\phi_1)$ or $\cos(n\phi_{-1})$. Equation (\ref{Gapping}) is the gapping condition for the perturbations. For instance, the pairing perturbation discussed in section \ref{sec:gapless} satisfies the gapping condition. On the other hand, if this condition is not satisfied for some perturbation, for example,  a Zeeman field coupling 
\[
H'=B_xS_x\sim \cos(2\phi_1 +\phi_{-1}) +\cos(2\phi_{-1}+\phi_1)
\] 
which does not contain the Higgs term, then the boundary remains gapless even the symmetry is explicitly broken. This is verified by our numerical result shown in Fig.~\ref{fig:app}, where the correlation function $\langle Q^x(r)Q^x(r+x)\rangle$ remains power law if we add a Zeeman field $B_x=0.4$ (in units of $t_{ij}$) to the whole system.

Finally, we give the Chern-Simons theory of the CLS state where $\mathcal C_1=\mathcal C_0=\mathcal C_{-1}=1$. Form (\ref{CS2}), the $K$ matrix of the CSL is given as $K=\left(\begin{matrix}2&1\\1&2\end{matrix}\right)$. Since $\det K=3$, The ground state degeneracy of CSL on a torus is 3. Furthermore, since the gapping condition (\ref{Gapping}) has no solutions, the boundary can not be gapped out by small local perturbations. 

%%%%%%%%%%%%%%%%%%%%%%%%%%%%%%%%%%%%
\bibliography{Liuzx}

\begin{thebibliography}{55}%
\makeatletter
\providecommand \@ifxundefined [1]{%
 \@ifx{#1\undefined}
}%
\providecommand \@ifnum [1]{%
 \ifnum #1\expandafter \@firstoftwo
 \else \expandafter \@secondoftwo
 \fi
}%
\providecommand \@ifx [1]{%
 \ifx #1\expandafter \@firstoftwo
 \else \expandafter \@secondoftwo
 \fi
}%
\providecommand \natexlab [1]{#1}%
\providecommand \enquote  [1]{``#1''}%
\providecommand \bibnamefont  [1]{#1}%
\providecommand \bibfnamefont [1]{#1}%
\providecommand \citenamefont [1]{#1}%
\providecommand \href@noop [0]{\@secondoftwo}%
\providecommand \href [0]{\begingroup \@sanitize@url \@href}%
\providecommand \@href[1]{\@@startlink{#1}\@@href}%
\providecommand \@@href[1]{\endgroup#1\@@endlink}%
\providecommand \@sanitize@url [0]{\catcode `\\12\catcode `\$12\catcode
  `\&12\catcode `\#12\catcode `\^12\catcode `\_12\catcode `\%12\relax}%
\providecommand \@@startlink[1]{}%
\providecommand \@@endlink[0]{}%
\providecommand \url  [0]{\begingroup\@sanitize@url \@url }%
\providecommand \@url [1]{\endgroup\@href {#1}{\urlprefix }}%
\providecommand \urlprefix  [0]{URL }%
\providecommand \Eprint [0]{\href }%
\providecommand \doibase [0]{http://dx.doi.org/}%
\providecommand \selectlanguage [0]{\@gobble}%
\providecommand \bibinfo  [0]{\@secondoftwo}%
\providecommand \bibfield  [0]{\@secondoftwo}%
\providecommand \translation [1]{[#1]}%
\providecommand \BibitemOpen [0]{}%
\providecommand \bibitemStop [0]{}%
\providecommand \bibitemNoStop [0]{.\EOS\space}%
\providecommand \EOS [0]{\spacefactor3000\relax}%
\providecommand \BibitemShut  [1]{\csname bibitem#1\endcsname}%
\let\auto@bib@innerbib\@empty
%</preamble>
\bibitem [{\citenamefont {Wen}(1989)}]{Wen89}%
  \BibitemOpen
  \bibfield  {author} {\bibinfo {author} {\bibfnamefont {X.~G.}\ \bibnamefont
  {Wen}},\ }\href {\doibase 10.1103/PhysRevB.40.7387} {\bibfield  {journal}
  {\bibinfo  {journal} {Phys. Rev. B}\ }\textbf {\bibinfo {volume} {40}},\
  \bibinfo {pages} {7387} (\bibinfo {year} {1989})}\BibitemShut {NoStop}%
\bibitem [{\citenamefont {Wen}\ and\ \citenamefont {Niu}(1990)}]{WenNiu90}%
  \BibitemOpen
  \bibfield  {author} {\bibinfo {author} {\bibfnamefont {X.~G.}\ \bibnamefont
  {Wen}}\ and\ \bibinfo {author} {\bibfnamefont {Q.}~\bibnamefont {Niu}},\
  }\href {\doibase 10.1103/PhysRevB.41.9377} {\bibfield  {journal} {\bibinfo
  {journal} {Phys. Rev. B}\ }\textbf {\bibinfo {volume} {41}},\ \bibinfo
  {pages} {9377} (\bibinfo {year} {1990})}\BibitemShut {NoStop}%
\bibitem [{\citenamefont {Wen}(1990)}]{Wen1990}%
  \BibitemOpen
  \bibfield  {author} {\bibinfo {author} {\bibfnamefont {X.~G.}\ \bibnamefont
  {Wen}},\ }\href {\doibase 10.1142/S0217979290000139} {\bibfield  {journal}
  {\bibinfo  {journal} {International Journal of Modern Physics B}\ }\textbf
  {\bibinfo {volume} {04}},\ \bibinfo {pages} {239} (\bibinfo {year}
  {1990})}\BibitemShut {NoStop}%
\bibitem [{\citenamefont {Tsui}\ \emph {et~al.}(1982)\citenamefont {Tsui},
  \citenamefont {Stormer},\ and\ \citenamefont
  {Gossard}}]{TsuiStormerGossard1982}%
  \BibitemOpen
  \bibfield  {author} {\bibinfo {author} {\bibfnamefont {D.~C.}\ \bibnamefont
  {Tsui}}, \bibinfo {author} {\bibfnamefont {H.~L.}\ \bibnamefont {Stormer}}, \
  and\ \bibinfo {author} {\bibfnamefont {A.~C.}\ \bibnamefont {Gossard}},\
  }\href {\doibase 10.1103/PhysRevLett.48.1559} {\bibfield  {journal} {\bibinfo
   {journal} {Phys. Rev. Lett.}\ }\textbf {\bibinfo {volume} {48}},\ \bibinfo
  {pages} {1559} (\bibinfo {year} {1982})}\BibitemShut {NoStop}%
\bibitem [{\citenamefont {Laughlin}(1983)}]{Laughlin1983}%
  \BibitemOpen
  \bibfield  {author} {\bibinfo {author} {\bibfnamefont {R.~B.}\ \bibnamefont
  {Laughlin}},\ }\href {\doibase 10.1103/PhysRevLett.50.1395} {\bibfield
  {journal} {\bibinfo  {journal} {Phys. Rev. Lett.}\ }\textbf {\bibinfo
  {volume} {50}},\ \bibinfo {pages} {1395} (\bibinfo {year}
  {1983})}\BibitemShut {NoStop}%
\bibitem [{\citenamefont {Anderson}(1973)}]{Anderson1973}%
  \BibitemOpen
  \bibfield  {author} {\bibinfo {author} {\bibfnamefont {P.}~\bibnamefont
  {Anderson}},\ }\href {\doibase 10.1016/0025-5408(73)90167-0} {\bibfield
  {journal} {\bibinfo  {journal} {Materials Research Bulletin}\ }\textbf
  {\bibinfo {volume} {8}},\ \bibinfo {pages} {153 } (\bibinfo {year}
  {1973})}\BibitemShut {NoStop}%
\bibitem [{\citenamefont {{Anderson}}(1987)}]{Anderson1987}%
  \BibitemOpen
  \bibfield  {author} {\bibinfo {author} {\bibfnamefont {P.~W.}\ \bibnamefont
  {{Anderson}}},\ }\href {\doibase 10.1126/science.235.4793.1196} {\bibfield
  {journal} {\bibinfo  {journal} {Science}\ }\textbf {\bibinfo {volume}
  {235}},\ \bibinfo {pages} {1196} (\bibinfo {year} {1987})}\BibitemShut
  {NoStop}%
\bibitem [{\citenamefont {Landau}(1937)}]{Landau37}%
  \BibitemOpen
  \bibfield  {author} {\bibinfo {author} {\bibfnamefont {L.~D.}\ \bibnamefont
  {Landau}},\ }\href@noop {} {\bibfield  {journal} {\bibinfo  {journal} {Phys.
  Z. Sowjetunion}\ }\textbf {\bibinfo {volume} {11}},\ \bibinfo {pages} {26}
  (\bibinfo {year} {1937})}\BibitemShut {NoStop}%
\bibitem [{\citenamefont {Ginzburg}\ and\ \citenamefont
  {Landau}(1950)}]{GL5064}%
  \BibitemOpen
  \bibfield  {author} {\bibinfo {author} {\bibfnamefont {V.~L.}\ \bibnamefont
  {Ginzburg}}\ and\ \bibinfo {author} {\bibfnamefont {L.~D.}\ \bibnamefont
  {Landau}},\ }\href@noop {} {\bibfield  {journal} {\bibinfo  {journal} {Zh.
  Eksp. Teor. Fiz.}\ }\textbf {\bibinfo {volume} {20}},\ \bibinfo {pages}
  {1064} (\bibinfo {year} {1950})}\BibitemShut {NoStop}%
\bibitem [{\citenamefont {Chen}\ \emph {et~al.}(2010)\citenamefont {Chen},
  \citenamefont {Gu},\ and\ \citenamefont {Wen}}]{ChenGuWen2010}%
  \BibitemOpen
  \bibfield  {author} {\bibinfo {author} {\bibfnamefont {X.}~\bibnamefont
  {Chen}}, \bibinfo {author} {\bibfnamefont {Z.-C.}\ \bibnamefont {Gu}}, \ and\
  \bibinfo {author} {\bibfnamefont {X.-G.}\ \bibnamefont {Wen}},\ }\href
  {\doibase 10.1103/PhysRevB.82.155138} {\bibfield  {journal} {\bibinfo
  {journal} {Phys. Rev. B}\ }\textbf {\bibinfo {volume} {82}},\ \bibinfo
  {pages} {155138} (\bibinfo {year} {2010})}\BibitemShut {NoStop}%
\bibitem [{\citenamefont {Kitaev}\ and\ \citenamefont
  {Preskill}(2006)}]{PreskillKitaev06_TEE}%
  \BibitemOpen
  \bibfield  {author} {\bibinfo {author} {\bibfnamefont {A.}~\bibnamefont
  {Kitaev}}\ and\ \bibinfo {author} {\bibfnamefont {J.}~\bibnamefont
  {Preskill}},\ }\href {\doibase 10.1103/PhysRevLett.96.110404} {\bibfield
  {journal} {\bibinfo  {journal} {Phys. Rev. Lett.}\ }\textbf {\bibinfo
  {volume} {96}},\ \bibinfo {pages} {110404} (\bibinfo {year}
  {2006})}\BibitemShut {NoStop}%
\bibitem [{\citenamefont {Levin}\ and\ \citenamefont
  {Wen}(2006)}]{LevinWen06_TEE}%
  \BibitemOpen
  \bibfield  {author} {\bibinfo {author} {\bibfnamefont {M.}~\bibnamefont
  {Levin}}\ and\ \bibinfo {author} {\bibfnamefont {X.-G.}\ \bibnamefont
  {Wen}},\ }\href {\doibase 10.1103/PhysRevLett.96.110405} {\bibfield
  {journal} {\bibinfo  {journal} {Phys. Rev. Lett.}\ }\textbf {\bibinfo
  {volume} {96}},\ \bibinfo {pages} {110405} (\bibinfo {year}
  {2006})}\BibitemShut {NoStop}%
\bibitem [{\citenamefont {Gu}\ and\ \citenamefont {Wen}(2009)}]{GuWen2009}%
  \BibitemOpen
  \bibfield  {author} {\bibinfo {author} {\bibfnamefont {Z.-C.}\ \bibnamefont
  {Gu}}\ and\ \bibinfo {author} {\bibfnamefont {X.-G.}\ \bibnamefont {Wen}},\
  }\href {\doibase 10.1103/PhysRevB.80.155131} {\bibfield  {journal} {\bibinfo
  {journal} {Phys. Rev. B}\ }\textbf {\bibinfo {volume} {80}},\ \bibinfo
  {pages} {155131} (\bibinfo {year} {2009})}\BibitemShut {NoStop}%
\bibitem [{\citenamefont {Pollmann}\ \emph {et~al.}(2010)\citenamefont
  {Pollmann}, \citenamefont {Turner}, \citenamefont {Berg},\ and\ \citenamefont
  {Oshikawa}}]{Pollmann2010}%
  \BibitemOpen
  \bibfield  {author} {\bibinfo {author} {\bibfnamefont {F.}~\bibnamefont
  {Pollmann}}, \bibinfo {author} {\bibfnamefont {A.~M.}\ \bibnamefont
  {Turner}}, \bibinfo {author} {\bibfnamefont {E.}~\bibnamefont {Berg}}, \ and\
  \bibinfo {author} {\bibfnamefont {M.}~\bibnamefont {Oshikawa}},\ }\href
  {\doibase 10.1103/PhysRevB.81.064439} {\bibfield  {journal} {\bibinfo
  {journal} {Phys. Rev. B}\ }\textbf {\bibinfo {volume} {81}},\ \bibinfo
  {pages} {064439} (\bibinfo {year} {2010})}\BibitemShut {NoStop}%
\bibitem [{\citenamefont {Haldane}(1983)}]{HaldanePLA1983}%
  \BibitemOpen
  \bibfield  {author} {\bibinfo {author} {\bibfnamefont {F.}~\bibnamefont
  {Haldane}},\ }\href {\doibase 10.1016/0375-9601(83)90631-X} {\bibfield
  {journal} {\bibinfo  {journal} {Physics Letters A}\ }\textbf {\bibinfo
  {volume} {93}},\ \bibinfo {pages} {464 } (\bibinfo {year}
  {1983})}\BibitemShut {NoStop}%
\bibitem [{\citenamefont {{Haldane}}(1983)}]{HaldanePRL1983}%
  \BibitemOpen
  \bibfield  {author} {\bibinfo {author} {\bibfnamefont {F.~D.~M.}\
  \bibnamefont {{Haldane}}},\ }\href {\doibase 10.1103/PhysRevLett.50.1153}
  {\bibfield  {journal} {\bibinfo  {journal} {Physical Review Letters}\
  }\textbf {\bibinfo {volume} {50}},\ \bibinfo {pages} {1153} (\bibinfo {year}
  {1983})}\BibitemShut {NoStop}%
\bibitem [{\citenamefont {Kane}\ and\ \citenamefont {Mele}(2005)}]{KM0502}%
  \BibitemOpen
  \bibfield  {author} {\bibinfo {author} {\bibfnamefont {C.~L.}\ \bibnamefont
  {Kane}}\ and\ \bibinfo {author} {\bibfnamefont {E.~J.}\ \bibnamefont
  {Mele}},\ }\href@noop {} {\bibfield  {journal} {\bibinfo  {journal} {Phys.
  Rev. Lett.}\ }\textbf {\bibinfo {volume} {95}},\ \bibinfo {pages} {146802}
  (\bibinfo {year} {2005})},\ \Eprint {http://arxiv.org/abs/cond-mat/0506581}
  {cond-mat/0506581} \BibitemShut {NoStop}%
\bibitem [{\citenamefont {Bernevig}\ and\ \citenamefont
  {Zhang}(2006)}]{BZ0602}%
  \BibitemOpen
  \bibfield  {author} {\bibinfo {author} {\bibfnamefont {B.~A.}\ \bibnamefont
  {Bernevig}}\ and\ \bibinfo {author} {\bibfnamefont {S.-C.}\ \bibnamefont
  {Zhang}},\ }\href@noop {} {\bibfield  {journal} {\bibinfo  {journal} {Phys.
  Rev. Lett.}\ }\textbf {\bibinfo {volume} {96}},\ \bibinfo {pages} {106802}
  (\bibinfo {year} {2006})}\BibitemShut {NoStop}%
\bibitem [{\citenamefont {Moore}\ and\ \citenamefont {Balents}(2007)}]{MB0706}%
  \BibitemOpen
  \bibfield  {author} {\bibinfo {author} {\bibfnamefont {J.~E.}\ \bibnamefont
  {Moore}}\ and\ \bibinfo {author} {\bibfnamefont {L.}~\bibnamefont
  {Balents}},\ }\href@noop {} {\bibfield  {journal} {\bibinfo  {journal} {Phys.
  Rev. B}\ }\textbf {\bibinfo {volume} {75}},\ \bibinfo {pages} {121306}
  (\bibinfo {year} {2007})},\ \Eprint {http://arxiv.org/abs/cond-mat/0607314}
  {cond-mat/0607314} \BibitemShut {NoStop}%
\bibitem [{\citenamefont {Fu}\ \emph {et~al.}(2007)\citenamefont {Fu},
  \citenamefont {Kane},\ and\ \citenamefont {Mele}}]{FKM0703}%
  \BibitemOpen
  \bibfield  {author} {\bibinfo {author} {\bibfnamefont {L.}~\bibnamefont
  {Fu}}, \bibinfo {author} {\bibfnamefont {C.~L.}\ \bibnamefont {Kane}}, \ and\
  \bibinfo {author} {\bibfnamefont {E.~J.}\ \bibnamefont {Mele}},\ }\href@noop
  {} {\bibfield  {journal} {\bibinfo  {journal} {Phys. Rev. Lett.}\ }\textbf
  {\bibinfo {volume} {98}},\ \bibinfo {pages} {106803} (\bibinfo {year}
  {2007})},\ \Eprint {http://arxiv.org/abs/cond-mat/0607699} {cond-mat/0607699}
  \BibitemShut {NoStop}%
\bibitem [{\citenamefont {Qi}\ \emph {et~al.}(2008)\citenamefont {Qi},
  \citenamefont {Hughes},\ and\ \citenamefont {Zhang}}]{QHZ0824}%
  \BibitemOpen
  \bibfield  {author} {\bibinfo {author} {\bibfnamefont {X.-L.}\ \bibnamefont
  {Qi}}, \bibinfo {author} {\bibfnamefont {T.}~\bibnamefont {Hughes}}, \ and\
  \bibinfo {author} {\bibfnamefont {S.-C.}\ \bibnamefont {Zhang}},\ }\href@noop
  {} {\bibfield  {journal} {\bibinfo  {journal} {Phys. Rev. B}\ }\textbf
  {\bibinfo {volume} {78}},\ \bibinfo {pages} {195424} (\bibinfo {year}
  {2008})},\ \Eprint {http://arxiv.org/abs/arXiv:0802.3537} {arXiv:0802.3537}
  \BibitemShut {NoStop}%
\bibitem [{\citenamefont {Chen}\ \emph {et~al.}(2013)\citenamefont {Chen},
  \citenamefont {Gu}, \citenamefont {Liu},\ and\ \citenamefont
  {Wen}}]{ChenGuLiuWen2011}%
  \BibitemOpen
  \bibfield  {author} {\bibinfo {author} {\bibfnamefont {X.}~\bibnamefont
  {Chen}}, \bibinfo {author} {\bibfnamefont {Z.-C.}\ \bibnamefont {Gu}},
  \bibinfo {author} {\bibfnamefont {Z.-X.}\ \bibnamefont {Liu}}, \ and\
  \bibinfo {author} {\bibfnamefont {X.-G.}\ \bibnamefont {Wen}},\ }\href
  {\doibase 10.1103/PhysRevB.87.155114} {\bibfield  {journal} {\bibinfo
  {journal} {Phys. Rev. B}\ }\textbf {\bibinfo {volume} {87}},\ \bibinfo
  {pages} {155114} (\bibinfo {year} {2013})}\BibitemShut {NoStop}%
\bibitem [{\citenamefont {Bi}\ \emph {et~al.}(2013)\citenamefont {Bi},
  \citenamefont {Rasmussen},\ and\ \citenamefont {Xu}}]{Xu13}%
  \BibitemOpen
  \bibfield  {author} {\bibinfo {author} {\bibfnamefont {Z.}~\bibnamefont
  {Bi}}, \bibinfo {author} {\bibfnamefont {A.}~\bibnamefont {Rasmussen}}, \
  and\ \bibinfo {author} {\bibfnamefont {C.}~\bibnamefont {Xu}},\ }\href
  {http://arxiv.org/abs/arXiv:1309.0515} {\bibfield  {journal} {\bibinfo
  {journal} {arXiv:1309.0515}\ } (\bibinfo {year} {2013})}\BibitemShut
  {NoStop}%
\bibitem [{\citenamefont {Lu}\ and\ \citenamefont
  {Vishwanath}(2012)}]{LuVishwanath2012}%
  \BibitemOpen
  \bibfield  {author} {\bibinfo {author} {\bibfnamefont {Y.-M.}\ \bibnamefont
  {Lu}}\ and\ \bibinfo {author} {\bibfnamefont {A.}~\bibnamefont
  {Vishwanath}},\ }\href {\doibase 10.1103/PhysRevB.86.125119} {\bibfield
  {journal} {\bibinfo  {journal} {Phys. Rev. B}\ }\textbf {\bibinfo {volume}
  {86}},\ \bibinfo {pages} {125119} (\bibinfo {year} {2012})}\BibitemShut
  {NoStop}%
\bibitem [{\citenamefont {Chen}\ and\ \citenamefont {Wen}(2012)}]{ChenWen2012}%
  \BibitemOpen
  \bibfield  {author} {\bibinfo {author} {\bibfnamefont {X.}~\bibnamefont
  {Chen}}\ and\ \bibinfo {author} {\bibfnamefont {X.-G.}\ \bibnamefont {Wen}},\
  }\href {\doibase 10.1103/PhysRevB.86.235135} {\bibfield  {journal} {\bibinfo
  {journal} {Phys. Rev. B}\ }\textbf {\bibinfo {volume} {86}},\ \bibinfo
  {pages} {235135} (\bibinfo {year} {2012})}\BibitemShut {NoStop}%
\bibitem [{\citenamefont {Senthil}\ and\ \citenamefont
  {Levin}(2013)}]{SenthilLevin13}%
  \BibitemOpen
  \bibfield  {author} {\bibinfo {author} {\bibfnamefont {T.}~\bibnamefont
  {Senthil}}\ and\ \bibinfo {author} {\bibfnamefont {M.}~\bibnamefont
  {Levin}},\ }\href {\doibase 10.1103/PhysRevLett.110.046801} {\bibfield
  {journal} {\bibinfo  {journal} {Phys. Rev. Lett.}\ }\textbf {\bibinfo
  {volume} {110}},\ \bibinfo {pages} {046801} (\bibinfo {year}
  {2013})}\BibitemShut {NoStop}%
\bibitem [{\citenamefont {Regnault}\ and\ \citenamefont
  {Senthil}(2013)}]{RegnaultSenthil13}%
  \BibitemOpen
  \bibfield  {author} {\bibinfo {author} {\bibfnamefont {N.}~\bibnamefont
  {Regnault}}\ and\ \bibinfo {author} {\bibfnamefont {T.}~\bibnamefont
  {Senthil}},\ }\href {\doibase 10.1103/PhysRevB.88.161106} {\bibfield
  {journal} {\bibinfo  {journal} {Phys. Rev. B}\ }\textbf {\bibinfo {volume}
  {88}},\ \bibinfo {pages} {161106} (\bibinfo {year} {2013})}\BibitemShut
  {NoStop}%
\bibitem [{\citenamefont {Liu}\ and\ \citenamefont {Wen}(2013)}]{LiuWen2012}%
  \BibitemOpen
  \bibfield  {author} {\bibinfo {author} {\bibfnamefont {Z.-X.}\ \bibnamefont
  {Liu}}\ and\ \bibinfo {author} {\bibfnamefont {X.-G.}\ \bibnamefont {Wen}},\
  }\href {\doibase 10.1103/PhysRevLett.110.067205} {\bibfield  {journal}
  {\bibinfo  {journal} {Phys. Rev. Lett.}\ }\textbf {\bibinfo {volume} {110}},\
  \bibinfo {pages} {067205} (\bibinfo {year} {2013})}\BibitemShut {NoStop}%
\bibitem [{\citenamefont {Wen}(2002)}]{W0213}%
  \BibitemOpen
  \bibfield  {author} {\bibinfo {author} {\bibfnamefont {X.-G.}\ \bibnamefont
  {Wen}},\ }\href@noop {} {\bibfield  {journal} {\bibinfo  {journal} {Phys.
  Rev. B}\ }\textbf {\bibinfo {volume} {65}},\ \bibinfo {pages} {165113}
  (\bibinfo {year} {2002})},\ \Eprint {http://arxiv.org/abs/cond-mat/0107071}
  {cond-mat/0107071} \BibitemShut {NoStop}%
\bibitem [{\citenamefont {Kou}\ \emph {et~al.}(2008)\citenamefont {Kou},
  \citenamefont {Levin},\ and\ \citenamefont {Wen}}]{KLW0834}%
  \BibitemOpen
  \bibfield  {author} {\bibinfo {author} {\bibfnamefont {S.-P.}\ \bibnamefont
  {Kou}}, \bibinfo {author} {\bibfnamefont {M.}~\bibnamefont {Levin}}, \ and\
  \bibinfo {author} {\bibfnamefont {X.-G.}\ \bibnamefont {Wen}},\ }\href@noop
  {} {\bibfield  {journal} {\bibinfo  {journal} {Phys. Rev. B}\ }\textbf
  {\bibinfo {volume} {78}},\ \bibinfo {pages} {155134} (\bibinfo {year}
  {2008})},\ \Eprint {http://arxiv.org/abs/arXiv:0803.2300} {arXiv:0803.2300}
  \BibitemShut {NoStop}%
\bibitem [{\citenamefont {Kou}\ and\ \citenamefont {Wen}(2009)}]{KW0906}%
  \BibitemOpen
  \bibfield  {author} {\bibinfo {author} {\bibfnamefont {S.-P.}\ \bibnamefont
  {Kou}}\ and\ \bibinfo {author} {\bibfnamefont {X.-G.}\ \bibnamefont {Wen}},\
  }\href@noop {} {\bibfield  {journal} {\bibinfo  {journal} {Phys. Rev. B}\
  }\textbf {\bibinfo {volume} {80}},\ \bibinfo {pages} {224406} (\bibinfo
  {year} {2009})},\ \Eprint {http://arxiv.org/abs/arXiv:0907.4537}
  {arXiv:0907.4537} \BibitemShut {NoStop}%
\bibitem [{\citenamefont {Mesaros}\ and\ \citenamefont {Ran}(2013)}]{MR1315}%
  \BibitemOpen
  \bibfield  {author} {\bibinfo {author} {\bibfnamefont {A.}~\bibnamefont
  {Mesaros}}\ and\ \bibinfo {author} {\bibfnamefont {Y.}~\bibnamefont {Ran}},\
  }\href@noop {} {\bibfield  {journal} {\bibinfo  {journal} {Phys. Rev. B}\
  }\textbf {\bibinfo {volume} {87}},\ \bibinfo {pages} {155115} (\bibinfo
  {year} {2013})},\ \Eprint {http://arxiv.org/abs/arXiv:1212.0835}
  {arXiv:1212.0835} \BibitemShut {NoStop}%
\bibitem [{\citenamefont {Hung}\ and\ \citenamefont {Wan}(2013)}]{HungWan2013}%
  \BibitemOpen
  \bibfield  {author} {\bibinfo {author} {\bibfnamefont {L.-Y.}\ \bibnamefont
  {Hung}}\ and\ \bibinfo {author} {\bibfnamefont {Y.}~\bibnamefont {Wan}},\
  }\href {\doibase 10.1103/PhysRevB.87.195103} {\bibfield  {journal} {\bibinfo
  {journal} {Phys. Rev. B}\ }\textbf {\bibinfo {volume} {87}},\ \bibinfo
  {pages} {195103} (\bibinfo {year} {2013})}\BibitemShut {NoStop}%
\bibitem [{\citenamefont {Lu}\ and\ \citenamefont
  {Vishwanath}(2013)}]{LuVishwanath2013}%
  \BibitemOpen
  \bibfield  {author} {\bibinfo {author} {\bibfnamefont {Y.-M.}\ \bibnamefont
  {Lu}}\ and\ \bibinfo {author} {\bibfnamefont {A.}~\bibnamefont
  {Vishwanath}},\ }\href {http://arxiv.org/abs/arXiv:1302.2634} {\bibfield
  {journal} {\bibinfo  {journal} {arXiv:1302.2634}\ } (\bibinfo {year}
  {2013})}\BibitemShut {NoStop}%
\bibitem [{\citenamefont {Chen}\ \emph {et~al.}(2011)\citenamefont {Chen},
  \citenamefont {Liu},\ and\ \citenamefont {Wen}}]{ChenLiuWen2011}%
  \BibitemOpen
  \bibfield  {author} {\bibinfo {author} {\bibfnamefont {X.}~\bibnamefont
  {Chen}}, \bibinfo {author} {\bibfnamefont {Z.-X.}\ \bibnamefont {Liu}}, \
  and\ \bibinfo {author} {\bibfnamefont {X.-G.}\ \bibnamefont {Wen}},\ }\href
  {\doibase 10.1103/PhysRevB.84.235141} {\bibfield  {journal} {\bibinfo
  {journal} {Phys. Rev. B}\ }\textbf {\bibinfo {volume} {84}},\ \bibinfo
  {pages} {235141} (\bibinfo {year} {2011})}\BibitemShut {NoStop}%
\bibitem [{\citenamefont {Levin}\ and\ \citenamefont {Gu}(2012)}]{LevinGu2012}%
  \BibitemOpen
  \bibfield  {author} {\bibinfo {author} {\bibfnamefont {M.}~\bibnamefont
  {Levin}}\ and\ \bibinfo {author} {\bibfnamefont {Z.-C.}\ \bibnamefont {Gu}},\
  }\href {\doibase 10.1103/PhysRevB.86.115109} {\bibfield  {journal} {\bibinfo
  {journal} {Phys. Rev. B}\ }\textbf {\bibinfo {volume} {86}},\ \bibinfo
  {pages} {115109} (\bibinfo {year} {2012})}\BibitemShut {NoStop}%
\bibitem [{\citenamefont {Lu}\ and\ \citenamefont {Lee}(2014)}]{LuLee12}%
  \BibitemOpen
  \bibfield  {author} {\bibinfo {author} {\bibfnamefont {Y.-M.}\ \bibnamefont
  {Lu}}\ and\ \bibinfo {author} {\bibfnamefont {D.-H.}\ \bibnamefont {Lee}},\
  }\href {\doibase 10.1103/PhysRevB.89.184417} {\bibfield  {journal} {\bibinfo
  {journal} {Phys. Rev. B}\ }\textbf {\bibinfo {volume} {89}},\ \bibinfo
  {pages} {184417} (\bibinfo {year} {2014})}\BibitemShut {NoStop}%
\bibitem [{\citenamefont {Ye}\ and\ \citenamefont {Wen}(2013)}]{YeWen13}%
  \BibitemOpen
  \bibfield  {author} {\bibinfo {author} {\bibfnamefont {P.}~\bibnamefont
  {Ye}}\ and\ \bibinfo {author} {\bibfnamefont {X.-G.}\ \bibnamefont {Wen}},\
  }\href {\doibase 10.1103/PhysRevB.87.195128} {\bibfield  {journal} {\bibinfo
  {journal} {Phys. Rev. B}\ }\textbf {\bibinfo {volume} {87}},\ \bibinfo
  {pages} {195128} (\bibinfo {year} {2013})}\BibitemShut {NoStop}%
\bibitem [{\citenamefont {Liu}\ \emph {et~al.}(2013)\citenamefont {Liu},
  \citenamefont {Gu},\ and\ \citenamefont {Wen}}]{LiuGuWen14}%
  \BibitemOpen
  \bibfield  {author} {\bibinfo {author} {\bibfnamefont {Z.-X.}\ \bibnamefont
  {Liu}}, \bibinfo {author} {\bibfnamefont {Z.-C.}\ \bibnamefont {Gu}}, \ and\
  \bibinfo {author} {\bibfnamefont {X.-G.}\ \bibnamefont {Wen}},\ }\href
  {http://arxiv.org/abs/arXiv:1404.2818} {\bibfield  {journal} {\bibinfo
  {journal} {arXiv:1404.2818}\ } (\bibinfo {year} {2013})}\BibitemShut
  {NoStop}%
\bibitem [{\citenamefont {Mei}\ and\ \citenamefont {Wen}(2014)}]{MeiWen14}%
  \BibitemOpen
  \bibfield  {author} {\bibinfo {author} {\bibfnamefont {J.-W.}\ \bibnamefont
  {Mei}}\ and\ \bibinfo {author} {\bibfnamefont {X.-G.}\ \bibnamefont {Wen}},\
  }\href {http://arxiv.org/abs/ arXiv:1407.0869} {\bibfield  {journal}
  {\bibinfo  {journal} {arXiv:1407.0869}\ } (\bibinfo {year}
  {2014})}\BibitemShut {NoStop}%
\bibitem [{\citenamefont {Geraedts}\ and\ \citenamefont
  {Motrunich}(2014)}]{Motrunich14}%
  \BibitemOpen
  \bibfield  {author} {\bibinfo {author} {\bibfnamefont {S.~D.}\ \bibnamefont
  {Geraedts}}\ and\ \bibinfo {author} {\bibfnamefont {O.~I.}\ \bibnamefont
  {Motrunich}},\ }\href {http://arxiv.org/abs/arXiv:1408.1096} {\bibfield
  {journal} {\bibinfo  {journal} {arXiv:1408.1096}\ } (\bibinfo {year}
  {2014})}\BibitemShut {NoStop}%
\bibitem [{\citenamefont {Tu}\ \emph {et~al.}(2014)\citenamefont {Tu},
  \citenamefont {Nielsen},\ and\ \citenamefont {Sierra}}]{Tu14SUN_1}%
  \BibitemOpen
  \bibfield  {author} {\bibinfo {author} {\bibfnamefont {H.-H.}\ \bibnamefont
  {Tu}}, \bibinfo {author} {\bibfnamefont {A.~E.}\ \bibnamefont {Nielsen}}, \
  and\ \bibinfo {author} {\bibfnamefont {G.}~\bibnamefont {Sierra}},\ }\href
  {\doibase http://dx.doi.org/10.1016/j.nuclphysb.2014.06.027} {\bibfield
  {journal} {\bibinfo  {journal} {Nuclear Physics B}\ }\textbf {\bibinfo
  {volume} {886}},\ \bibinfo {pages} {328 } (\bibinfo {year}
  {2014})}\BibitemShut {NoStop}%
\bibitem [{\citenamefont {Lee}\ \emph {et~al.}(2006)\citenamefont {Lee},
  \citenamefont {Nagaosa},\ and\ \citenamefont {Wen}}]{PALee06RMP}%
  \BibitemOpen
  \bibfield  {author} {\bibinfo {author} {\bibfnamefont {P.}~\bibnamefont
  {Lee}}, \bibinfo {author} {\bibfnamefont {N.}~\bibnamefont {Nagaosa}}, \ and\
  \bibinfo {author} {\bibfnamefont {X.-G.}\ \bibnamefont {Wen}},\ }\href
  {\doibase 10.1103/RevModPhys.78.17} {\bibfield  {journal} {\bibinfo
  {journal} {Rev. Mod. Phys.}\ }\textbf {\bibinfo {volume} {78}},\ \bibinfo
  {pages} {17} (\bibinfo {year} {2006})}\BibitemShut {NoStop}%
\bibitem [{\citenamefont {Liu}\ \emph {et~al.}(2010{\natexlab{a}})\citenamefont
  {Liu}, \citenamefont {Zhou},\ and\ \citenamefont
  {Ng}}]{LiuZhouNg2011_FermionMF}%
  \BibitemOpen
  \bibfield  {author} {\bibinfo {author} {\bibfnamefont {Z.-X.}\ \bibnamefont
  {Liu}}, \bibinfo {author} {\bibfnamefont {Y.}~\bibnamefont {Zhou}}, \ and\
  \bibinfo {author} {\bibfnamefont {T.-K.}\ \bibnamefont {Ng}},\ }\href
  {\doibase 10.1103/PhysRevB.82.144422} {\bibfield  {journal} {\bibinfo
  {journal} {Phys. Rev. B}\ }\textbf {\bibinfo {volume} {82}},\ \bibinfo
  {pages} {144422} (\bibinfo {year} {2010}{\natexlab{a}})}\BibitemShut
  {NoStop}%
\bibitem [{\citenamefont {Liu}\ \emph {et~al.}(2010{\natexlab{b}})\citenamefont
  {Liu}, \citenamefont {Zhou},\ and\ \citenamefont
  {Ng}}]{LiuZhouNg2011_Spin-1SL}%
  \BibitemOpen
  \bibfield  {author} {\bibinfo {author} {\bibfnamefont {Z.-X.}\ \bibnamefont
  {Liu}}, \bibinfo {author} {\bibfnamefont {Y.}~\bibnamefont {Zhou}}, \ and\
  \bibinfo {author} {\bibfnamefont {T.-K.}\ \bibnamefont {Ng}},\ }\href
  {\doibase 10.1103/PhysRevB.81.224417} {\bibfield  {journal} {\bibinfo
  {journal} {Phys. Rev. B}\ }\textbf {\bibinfo {volume} {81}},\ \bibinfo
  {pages} {224417} (\bibinfo {year} {2010}{\natexlab{b}})}\BibitemShut
  {NoStop}%
\bibitem [{\citenamefont {Liu}\ \emph {et~al.}(2012)\citenamefont {Liu},
  \citenamefont {Zhou}, \citenamefont {Tu}, \citenamefont {Wen},\ and\
  \citenamefont {Ng}}]{LiuZhouTuWenNg2012}%
  \BibitemOpen
  \bibfield  {author} {\bibinfo {author} {\bibfnamefont {Z.-X.}\ \bibnamefont
  {Liu}}, \bibinfo {author} {\bibfnamefont {Y.}~\bibnamefont {Zhou}}, \bibinfo
  {author} {\bibfnamefont {H.-H.}\ \bibnamefont {Tu}}, \bibinfo {author}
  {\bibfnamefont {X.-G.}\ \bibnamefont {Wen}}, \ and\ \bibinfo {author}
  {\bibfnamefont {T.-K.}\ \bibnamefont {Ng}},\ }\href {\doibase
  10.1103/PhysRevB.85.195144} {\bibfield  {journal} {\bibinfo  {journal} {Phys.
  Rev. B}\ }\textbf {\bibinfo {volume} {85}},\ \bibinfo {pages} {195144}
  (\bibinfo {year} {2012})}\BibitemShut {NoStop}%
\bibitem [{\citenamefont {Bieri}\ \emph {et~al.}(2012)\citenamefont {Bieri},
  \citenamefont {Serbyn}, \citenamefont {Senthil},\ and\ \citenamefont
  {Lee}}]{PALee12Gutz}%
  \BibitemOpen
  \bibfield  {author} {\bibinfo {author} {\bibfnamefont {S.}~\bibnamefont
  {Bieri}}, \bibinfo {author} {\bibfnamefont {M.}~\bibnamefont {Serbyn}},
  \bibinfo {author} {\bibfnamefont {T.}~\bibnamefont {Senthil}}, \ and\
  \bibinfo {author} {\bibfnamefont {P.~A.}\ \bibnamefont {Lee}},\ }\href
  {\doibase 10.1103/PhysRevB.86.224409} {\bibfield  {journal} {\bibinfo
  {journal} {Phys. Rev. B}\ }\textbf {\bibinfo {volume} {86}},\ \bibinfo
  {pages} {224409} (\bibinfo {year} {2012})}\BibitemShut {NoStop}%
\bibitem [{Note1()}]{Note1}%
  \BibitemOpen
  \bibinfo {note} {$\protect \mathaccentV {tilde}07EJ_0=0$ is nothing but the
  number constraint, which induces $\protect \mathaccentV
  {tilde}07EJ_i=0$.}\BibitemShut {Stop}%
\bibitem [{\citenamefont {Gros}(1989)}]{Gros88}%
  \BibitemOpen
  \bibfield  {author} {\bibinfo {author} {\bibfnamefont {C.}~\bibnamefont
  {Gros}},\ }\href {\doibase http://dx.doi.org/10.1016/0003-4916(89)90077-8}
  {\bibfield  {journal} {\bibinfo  {journal} {Annals of Physics}\ }\textbf
  {\bibinfo {volume} {189}},\ \bibinfo {pages} {53 } (\bibinfo {year}
  {1989})}\BibitemShut {NoStop}%
\bibitem [{\citenamefont {Niu}\ \emph {et~al.}(1985)\citenamefont {Niu},
  \citenamefont {Thouless},\ and\ \citenamefont {Wu}}]{NTDW_85Twisted}%
  \BibitemOpen
  \bibfield  {author} {\bibinfo {author} {\bibfnamefont {Q.}~\bibnamefont
  {Niu}}, \bibinfo {author} {\bibfnamefont {D.~J.}\ \bibnamefont {Thouless}}, \
  and\ \bibinfo {author} {\bibfnamefont {Y.-S.}\ \bibnamefont {Wu}},\ }\href
  {\doibase 10.1103/PhysRevB.31.3372} {\bibfield  {journal} {\bibinfo
  {journal} {Phys. Rev. B}\ }\textbf {\bibinfo {volume} {31}},\ \bibinfo
  {pages} {3372} (\bibinfo {year} {1985})}\BibitemShut {NoStop}%
\bibitem [{\citenamefont {Hastings}\ \emph {et~al.}(2010)\citenamefont
  {Hastings}, \citenamefont {Gonz\'alez}, \citenamefont {Kallin},\ and\
  \citenamefont {Melko}}]{Hastings_TEE}%
  \BibitemOpen
  \bibfield  {author} {\bibinfo {author} {\bibfnamefont {M.~B.}\ \bibnamefont
  {Hastings}}, \bibinfo {author} {\bibfnamefont {I.}~\bibnamefont
  {Gonz\'alez}}, \bibinfo {author} {\bibfnamefont {A.~B.}\ \bibnamefont
  {Kallin}}, \ and\ \bibinfo {author} {\bibfnamefont {R.~G.}\ \bibnamefont
  {Melko}},\ }\href {\doibase 10.1103/PhysRevLett.104.157201} {\bibfield
  {journal} {\bibinfo  {journal} {Phys. Rev. Lett.}\ }\textbf {\bibinfo
  {volume} {104}},\ \bibinfo {pages} {157201} (\bibinfo {year}
  {2010})}\BibitemShut {NoStop}%
\bibitem [{\citenamefont {Cirac}\ and\ \citenamefont
  {Sierra}(2010)}]{Cirac_CFT}%
  \BibitemOpen
  \bibfield  {author} {\bibinfo {author} {\bibfnamefont {J.~I.}\ \bibnamefont
  {Cirac}}\ and\ \bibinfo {author} {\bibfnamefont {G.}~\bibnamefont {Sierra}},\
  }\href {\doibase 10.1103/PhysRevB.81.104431} {\bibfield  {journal} {\bibinfo
  {journal} {Phys. Rev. B}\ }\textbf {\bibinfo {volume} {81}},\ \bibinfo
  {pages} {104431} (\bibinfo {year} {2010})}\BibitemShut {NoStop}%
\bibitem [{\citenamefont {Zhang}\ \emph
  {et~al.}(2011{\natexlab{a}})\citenamefont {Zhang}, \citenamefont {Grover},\
  and\ \citenamefont {Vishwanath}}]{Zhang2011PRL}%
  \BibitemOpen
  \bibfield  {author} {\bibinfo {author} {\bibfnamefont {Y.}~\bibnamefont
  {Zhang}}, \bibinfo {author} {\bibfnamefont {T.}~\bibnamefont {Grover}}, \
  and\ \bibinfo {author} {\bibfnamefont {A.}~\bibnamefont {Vishwanath}},\
  }\href {\doibase 10.1103/PhysRevLett.107.067202} {\bibfield  {journal}
  {\bibinfo  {journal} {Phys. Rev. Lett.}\ }\textbf {\bibinfo {volume} {107}},\
  \bibinfo {pages} {067202} (\bibinfo {year} {2011}{\natexlab{a}})}\BibitemShut
  {NoStop}%
\bibitem [{\citenamefont {Zhang}\ \emph
  {et~al.}(2011{\natexlab{b}})\citenamefont {Zhang}, \citenamefont {Grover},\
  and\ \citenamefont {Vishwanath}}]{Zhang2011PRB}%
  \BibitemOpen
  \bibfield  {author} {\bibinfo {author} {\bibfnamefont {Y.}~\bibnamefont
  {Zhang}}, \bibinfo {author} {\bibfnamefont {T.}~\bibnamefont {Grover}}, \
  and\ \bibinfo {author} {\bibfnamefont {A.}~\bibnamefont {Vishwanath}},\
  }\href {\doibase 10.1103/PhysRevB.84.075128} {\bibfield  {journal} {\bibinfo
  {journal} {Phys. Rev. B}\ }\textbf {\bibinfo {volume} {84}},\ \bibinfo
  {pages} {075128} (\bibinfo {year} {2011}{\natexlab{b}})}\BibitemShut
  {NoStop}%
\bibitem [{\citenamefont {Pei}\ \emph {et~al.}(2013)\citenamefont {Pei},
  \citenamefont {Han}, \citenamefont {Liao},\ and\ \citenamefont
  {Li}}]{PeiLi2013}%
  \BibitemOpen
  \bibfield  {author} {\bibinfo {author} {\bibfnamefont {J.}~\bibnamefont
  {Pei}}, \bibinfo {author} {\bibfnamefont {S.}~\bibnamefont {Han}}, \bibinfo
  {author} {\bibfnamefont {H.}~\bibnamefont {Liao}}, \ and\ \bibinfo {author}
  {\bibfnamefont {T.}~\bibnamefont {Li}},\ }\href {\doibase
  10.1103/PhysRevB.88.125135} {\bibfield  {journal} {\bibinfo  {journal} {Phys.
  Rev. B}\ }\textbf {\bibinfo {volume} {88}},\ \bibinfo {pages} {125135}
  (\bibinfo {year} {2013})}\BibitemShut {NoStop}%
\end{thebibliography}%

\end{document}